\documentclass[11pt]{article}
\usepackage[utf8]{inputenc}
\usepackage{color}
\usepackage{graphicx,amssymb,amsmath,empheq}
\usepackage{authblk}
\usepackage{amsthm}
\usepackage{empheq}
\usepackage{subfig} % caution: not compatible with revtex4
\usepackage{hyperref}
\hypersetup{colorlinks = true, linkcolor=black, citecolor=black, urlcolor=blue}
\usepackage{url}
\usepackage{cite}

\begin{document}

\title{Snowmass White Paper: New ideas for many-body quantum systems from string theory and black holes}
\author{Mike Blake$^{1}$, Yingfei Gu$^{2}$, Sean A. Hartnoll$^{3}$, Hong Liu$^{4}$, Andrew Lucas$^{5}$, Krishna Rajagopal$^{4}$, Brian Swingle$^{6}$, Beni Yoshida$^{7}$}
\affil{\it $^{1}$ School of Mathematics, University of Bristol,  Bristol BS8 1UG, UK \\
\it $^{2}$ California Institute of Technology, Pasadena, CA 91125, USA\\
\it $^{3}$ DAMTP, University of Cambridge, Cambridge CB3 0WA, UK \\
\it $^{4}$ CTP, Massachusetts Institute of Technology, Cambridge, MA 02139, USA \\
\it $^{5}$ Department of Physics and Center for Theory of Quantum Matter, \\
\it University of Colorado, Boulder CO 80309, USA\\
\it $^{6}$ Brandeis University, Waltham, MA 02453, USA\\
\it $^{7}$ Perimeter Institute for Theoretical Physics, Waterloo, Ontario N2L 2Y5, Canada
}
\date{}

\maketitle

\centerline{\bf Abstract}

During the last two decades many new insights into the dynamics of strongly coupled quantum many-body systems have been obtained using gauge/gravity duality, with black holes often playing a universal role. In this white paper we summarize the results obtained and offer some outlook for future developments, including the ongoing mutually beneficial feedback loop with the study of more general, not necessarily holographic, quantum many-body systems.

%\end{abstract}

\tableofcontents

\section{Introduction} 

The interpretation of black holes as strongly quantum mechanical many-body systems is rooted in the black hole thermodynamics developed many decades ago by Bekenstein, Hawking and others. A deeper understanding of this thermodynamics became possible following the discovery of gauge/gravity duality by Maldacena in 1997. By relating quantum gravity to certain `large $N$' quantum field theories without gravity,
the duality leads to a new description for strongly coupled many-body systems, with black holes playing a
%surprising and
universal role in characterizing the dynamics of the large $N$ theory at finite temperature and density. 

During the last two decades many new insights into the dynamics of strongly coupled quantum many-body systems have been obtained with the help of gauge/gravity duality. In the first decade, significant progress was made in understanding transport properties of both condensed matter systems and the quark-gluon plasma. Novel mechanisms for superconductivity and `strange metal' behavior were found, and an understanding was developed of the far-from-equilibrium dynamics of the quark-gluon plasma, including its creation in heavy ion collisions at the LHC and RHIC.

In the past decade, several different threads of research have coalesced to allow more fine grained probes of holographic many-body systems. Firstly, the theoretical confusion generated by various iterations of the black hole information paradox led to the application of probes such as out-of-time ordered correlation functions (OTOCs) and spectral statistics to holographic models. Secondly, the incorporation of quantum information theoretic ideas into holography --- originally via the Ryu-Takayangi conjecture for holographic entanglement entropy --- has grown to include powerful machinery such as quantum error correction, tensor networks and quantum complexity. Thirdly, the (re-)emergence of the SYK model as solvable large $N$ theory that incorporates aspects of both fully-fledged holography and realistic many-body systems has established an exciting new bridge between black holes and many-body systems. Futhermore, the interplay between holographic and weakly coupled modeling of phenomena in heavy ion collisions, serving to ``bracket reality'', and analyses of experimental data from the LHC and RHIC has broadened and deepened our understanding of these phenomena and of strongly coupled dynamics and hydrodynamization in general. 

In addition to providing new vistas on quantum many-body systems, the developments mentioned above have also deepened our understanding of black hole physics and the emergence of spacetime geometry itself on the gravity side of the duality, starting from a dual non-gravitational system.
%For examples, gravity calculations of various observables are sensitive to geometry inside the event horizon of a black hole, providing hints on its possible emergence in the boundary theory, a particularly puzzling question. 
In this white paper we summarize the insights obtained into quantum many-body dynamics and offer some outlook for future developments, including the ongoing mutually beneficial feedback loop with the study of more general, not necessarily holographic, quantum many-body systems.

\section{Black holes and hydrodynamic transport}
\label{sec:hydro}

After its discovery, gauge/gravity duality was rapidly understood to imply that the classical gravitational dynamics of black hole horizons is equivalent to the dissipative nonzero temperature physics of certain large $N$ quantum field theories. By the early
2000's, gravitational computations were being used to obtain results about thermalization times and linear response functions of the dual nonzero temperature medium. Results of this kind are nontrivial to obtain on any strongly quantum medium by conventional methods.

One highlight of early work sprung from the realization that the long wavelength response functions of conserved densities were governed by relativistic (conformal) hydrodynamics \cite{Policastro:2001yc, Kovtun:2003wp}.
These are the most universal of perturbations away from thermodynamic equilibrium, characterized by a small number of transport coefficients. It was discovered that all holographic models with Einstein gravity duals have a shear viscosity given by \cite{Kovtun:2004de}:
\begin{equation}
\eta = \frac{\hbar}{4\pi k_B} s. \label{eq:KSSviscosity}
\end{equation}
Here $s$ is the entropy density. The impact of this result is still being felt across multiple fields of theoretical physics. In particular, the ability to directly exhibit hydrodynamic behavior within a new class of model systems led to a broader realization that strongly coupled quantum fluids are excellent platforms to observe hydrodynamics in experiments on quantum matter. Over the past decade, hunts for viscosities with ``minimal" $\eta$ comparable to the prediction in (\ref{eq:KSSviscosity}) have occurred in quark-gluon plasma \cite{Shuryak:2008eq}, cold atoms \cite{Cao_2010}, and in charge-neutral graphene \cite{muller2009, Crossno1058,Ku:2019lgj}.  More generally, a very active field  of ``electron hydrodynamics" (both theoretical and experimental) has emerged in the past few years, largely inspired by the developments highlighted above \cite{Bandurin1055,Krishna_Kumar_2017, sulpizio}; see \cite{Lucas:2017idv} for a review. The quark-gluon plasma is discussed further in section \ref{sec:qgp} below.

In section \ref{sec:far} below we will discuss large perturbations that take the black hole far from equilibrium --- in this regime, too, the classic gravitational dynamics provides solvable models of intricate many-body behavior.

\section{Holographic superconductors and strange metals}

The 2000's and early 2010's saw many results on the phase structure and response functions of large $N$ many-body systems with holographic duals. This included the discovery of new instabilities of black holes towards superconducting \cite{Gubser:2008px, Hartnoll:2008vx, Hartnoll:2008kx} and spatially modulated \cite{Nakamura:2009tf, Donos:2011bh} phases. The resulting Holographic superconductors and superfluids offer examples of superconductivity in strongly coupled systems from a novel mechanisms distinct from the familiar BCS theory. They also provide laboratories for exploring far-from-equilibrium superfluid dynamics, as we will discuss in Sec.~\ref{sec:far}. 

Holographic systems furthermore offer examples of strange metals that share phenomenological features with the strange metal phase of unconventional materials including high temperature superconductors. The simplest examples are the extremely rich `semi-local' physics associated to the emergence of an AdS$_2 \times \mathbb{R}^d$ near-horizon geometry of low temperature black holes. Early works emphasizing the importance of this geometry for instabilities, correlation functions and transport include \cite{Denef:2009tp,Liu:2009dm,Cubrovic:2009ye,Faulkner:2009wj,faulkner2010strange,Iqbal:2011in, Hartnoll:2012rj}. Motivated by the desire to capture more general dynamical exponents and infrared scaling symmetries of strange metals, new classes of near-horizon scaling geometries were discovered that could break Lorentz invariance \cite{Kachru:2008yh, Son:2008ye, Balasubramanian:2008dm} and exhibited 
hyperscaling violation \cite{Charmousis:2010zz,Gouteraux:2011ce, Huijse:2011ef}. These results provided insights into the behavior of more general strongly quantum media arising in nuclear and condensed matter physics, and were collected in several books and reviews including \cite{Iqbal:2011ae,Casalderrey-Solana:2011dxg,Zaanen:2015oix, hartnoll2018holographic,Liu:2020rrn,Zaanen:2021llz}. In particular, the cited books were collaborations between high energy, nuclear and condensed matter theorists.

More recent `spin-offs' from holographic work on unconventional phases of matter include new renormalization group approaches to disordered systems \cite{Hartnoll:2014cua, Narovlansky:2018muj, Ganesan:2021gun} and a systematic understanding of dissipation by pseudo-Goldstone bosons \cite{Amoretti:2018tzw, Delacretaz:2021qqu, Armas:2021vku}. Furthermore, recent work has seen the construction of holographic doped Mott insulators that incorporate key aspects of realistic strange metal phases~\cite{Andrade:2017ghg}. The SYK model has also enabled the construction of phenomenological models of strange metals that closely connect with holographic ones, as will be discussed in Sec.~\ref{sec:SYK}. The similarity between the SYK and holographic models was first emphasized in \cite{Sachdev:2010um}.

%In the past decade, several different threads of research have coalesced to allow more fine grained probes of holographic many-body systems. Firstly, the theoretical confusion generated by various iterations of the black hole information paradox led to the application of probes such as out-of-time ordered correlation functions (OTOCs) and spectral statistics to holographic models. Secondly, the incorporation of quantum information theoretic ideas into holography --- originally via the Ryu-Takayangi conjecture for holographic entanglement entropy --- has grown to include powerful machinery such as quantum error correction, tensor networks and quantum complexity. Thirdly, the (re-)emergence of the SYK model as solvable large $N$ many-body system that incorporates aspects of both fully-fledged holography and realistic many-body systems has established an exciting new bridge between black holes and many-body systems.

%Several of the probes just listed are sensitive to geometry inside the event horizon, for example involving both components of the thermofield double in an essential way, and therefore access information that is not easily visible in conventional response functions. In the remainder of this white paper we will elaborate on these themes in more detail, emphasizing the ongoing mutually beneficial feedback loop with the study of more general, not necessarily holographic, quantum many-body systems.

\section{Quantum dynamics and the onset of chaos}
\label{sec:onset}

A key breakthrough of the past decade has been the recognition of out-of-time ordered correlation functions \cite{1969Larkin} as novel probes of chaos and scrambling, originally identified in the context of holographic quantum field theories \cite{Shenker:2013pqa, Roberts:2014isa, Shenker:2014cwa}. In such theories these correlation functions were found to exhibit a regime of exponential growth with a rate that provides a quantum analogue of a classical Lyapunov exponent, and to spread ballistically in space at a speed known as the butterfly velocity. Similar exponential growth was soon after identified in the SYK model \cite{Kit.KITP.2, sachdev1993, Kitaev:2017awl, Maldacena:2016hyu}, and a ballistic spreading of OTOCs has subsequently been observed in a wide range of chaotic many-body quantum systems including random quantum circuits, conformal field theory and spin systems (e.g. in \cite{Nahum:2017yvy, vonKeyserlingk:2017dyr, Khemani:2017nda, Swingle:2016jdj, Turiaci:2016cvo, Gu:2016oyy}). Particularly significant results to arise from this activity include the discovery of a fundamental bound of chaos in large $N$ quantum systems \cite{Maldacena:2015waa}, and novel connections between many-body quantum chaos, transport and hydrodynamics \cite{Blake:2016wvh, Blake:2016sud, Grozdanov:2017ajz, Blake:2017ris, Blake:2018leo}.  We will return to bound on chaos and transport in section \ref{sec:bounds} below and to the SYK model in section \ref{sec:SYK}.

Despite these exciting developments, the study of chaos in many-body quantum systems is an emerging field, with many fundamental questions yet to be addressed. For instance, existing studies of OTOCs primarily take place in the context of specific models. Nevertheless many universal aspects of scrambling, such as exponential growth of OTOCs at large $N$ and the ballistic spreading of operators, are common to a wide range of many-body systems and quantum field theories. A key open question is whether one can develop general techniques capable of describing scrambling across a wide range of systems. For maximally chaotic systems, i.e, those that saturate the chaos bound, there exist effective field theory descriptions of scrambling based on reparameterisation modes and quantum hydrodynamics \cite{Haehl:2018izb, Haehl:2019eae, Blake:2017ris, Blake:2021wqj}. However currently there do not exist any such approaches for generic large $N$ systems that do not saturate the chaos bound. Closely related to this question are results from studies of holographic systems and SYK models that suggest a fundamental difference between the nature of scrambling in maximally chaotic and non-maximally chaotic systems \cite{Kitaev:2017awl, Gu:2018jsv, Blake:2021wqj}. Can one, perhaps using effective field theory, understand the apparent physical distinction between scrambling in such systems? 

Furthermore, it currently remains unclear whether the growth of OTOCs can be viewed as a precise characterisation of `chaos' in many-body quantum systems. Ballistic spreading of OTOCs is also observed in examples of integrable many-body quantum systems, and it would be interesting to understand if there exist alternative probes that can more sharply distinguish between chaotic and integrable many-body systems. It would also be interesting to understand in what sense `early-time' quantum chaos as probed by OTOCs is related to more traditional notions of `late-time' quantum chaos studied by the mathematical physics community such as the statistics of energy eigenvalues \cite{Cotler:2016fpe, Saad:2018bqo, Winer:2020gdp}. A key question for future research is therefore to try to establish connections between these distinct notions of quantum chaos. We return to spectral statistics in section \ref{sec:spectral} below.

Finally, a closely related aspect of quantum dynamics concerns the evolution of entanglement in a thermalizing quantum system. It has recently been proposed that a coarse-grained description of entanglement dynamics and scrambling in many-body quantum systems can be obtained in terms of the dynamics of an `entanglement membrane' \cite{Zhou:2018myl, Jonay:2018yei, Zhou:2019pob}. Such a description can explicitly be constructed in random quantum circuits \cite{Zhou:2018myl} and an entanglement membrane description of operator and state entanglement has also been obtained for holographic theories dual to classical gravity starting from the Ryu-Takayanagi proposal \cite{Mezei:2018jco, Mezei:2019zyt}. The fact that such a description arises in such distinct systems is remarkable, and a crucial open question is to understand the generality of such a description. For instance, can an entanglement membrane description of entanglement and operator dynamics be derived for some general class of quantum field theories, perhaps starting from permutation degrees of freedom recently identified in \cite{Liu:2020jsv}? Furthermore, the interplay of quantum scrambling, the chaotic behaviour of OTOC and entanglement has been understood to underlie the emergence of certain traversable wormhole solutions in holographic theories \cite{Gao:2016bin, Gao:2018yzk}. The geometric picture of the wormholes offered by gravity also suggests potentially new avenues for quantum many-body teleportation and a new phenomenon called `regenesis' \cite{Gao:2018yzk}. Further research in this direction could therefore lead to new insights into quantum communication.

\section{Bounding finite temperature quantum dynamics}
\label{sec:bounds}

We have already mentioned the `chaos bound', discovered in 2015 by Maldacena, Shenker, and Stanford \cite{Maldacena:2015waa}.  They demonstrated that under mild assumptions (which appear to be linked to the eigenstate thermalization hypothesis \cite{Murthy:2019fgs}), \begin{equation}
\mathrm{tr}\left(\rho^{1/4} A(t) \rho^{1/4} B \rho^{1/4} A(t) \rho^{1/4} B\right) \sim 1 - \frac{1}{N} e^{\lambda t} + \cdots, \label{eq:otoc}
\end{equation}
where $\rho \propto e^{-\beta H}$ is the thermal density matrix and \begin{equation}
\lambda \le \frac{2\pi k_B T}{\hbar}. \label{eq:lambdabound}
\end{equation}
The details of the regularization of the so-called `out-of-time-ordered correlator' in (\ref{eq:otoc}) appear unimportant so long as $\rho$ is split into at least two pieces \cite{Tsuji:2017fxs,Romero-Bermudez:2019vej}.  The peculiar mathematical object in (\ref{eq:otoc}) is motivated in part by considering a holographic thought experiment, where a small amount of matter falls towards one side of the maximally extended Schwarzchild black hole, and the gravitational backreaction (a shift between the horizons of each half) grows exponentially with time.  At infinite temperature, OTOCs probe operator growth, which is loosely the number of fundamental building blocks (such as Pauli matrices or creation operators) needed to write down the Heisenberg-evolved operator $A(t)$ \cite{Shenker:2013pqa, Nahum:2017yvy}. 

An important open question is to find a microscopic, holography-independent mechanism for (\ref{eq:lambdabound}).  A natural place to start is to generalize operator size to finite temperature.  Multiple proposals exist for how to do this \cite{Lucas:2018wsc, Qi:2018bje, Mousatov:2019xmc, Lensky:2020ubw}, and it has been argued to relate to the radial momentum of infalling particles in holographic models \cite{Brown:2018kvn}.   One promising possibility deserving further attention is that size winding explains finite $T$ operator growth in models with holographic duals \cite{Brown:2019hmk, Nezami:2021yaq, Schuster:2021uvg}.  Advances in the mathematical theory of bounds on quantum dynamics and operator growth \cite{Han:2018bqy, Lucas:2019cxr, Tran:2020xpc} give hope that the coming decade might see a comprehensive resolution of this puzzle.

In a remarkable parallel development, it has been suggested that the `Planckian' time scale appearing in the chaos bound above may possibly bound many-body timescales $\tau$ more generally \cite{Zaanen2004, Zaanen:2018edk, Hartnoll:2021ydi}
\begin{equation}
\frac{1}{\tau} \lesssim \frac{1}{\tau_{\mathrm{Pl}}} \sim \frac{k_{\mathrm{B}}T}{\hbar} \,. \label{eq:planckianbound}
\end{equation}
Within condensed matter physics, interest in the Planckian timescale is tied to the long-discussed $T$-linear resistivity above the superconducting temperature in high-$T_c$ superconductors for many years.  The conjecture that (\ref{eq:planckianbound}) is responsible for this $T$-linear resistivity saturating some upper bound on transport \cite{Hartnoll:2014lpa}, along with experimental evidence hinting that this is indeed the case \cite{Bruin2013, Legros}, has led to a huge effort over the past decade to classify models where $T$-linear resistivity is robust \emph{and} can be traced to a Planckian bound on dynamics.  In holographic models, e.g.~\cite{Donos:2012js, Gouteraux:2014hca, Donos:2014uba, Baggioli:2016oqk}, the $T$-scaling of any transport coefficient such as the resistivity is sensitive to matter couplings in the bulk action.  There appears to be more success in finding Planckian transport bounds in models based on lattice Sachdev-Ye-Kitaev models \cite{balents17, Patel:2017mjv, Chowdhury:2018sho, Patel:2019qce}, although $T$-linear resistivity can remain sensitive to certain model choices.

It is tantalizing to speculate that the two Planckian bounds: one on chaos, and one on transport, are intricately related.  The most well-explored idea \cite{Blake:2016wvh, Blake:2016sud, Gu:2016oyy, Baggioli:2016pia, Guo:2019csw} is that diffusion constants
\begin{equation}
D \sim \frac{v_B^2}{\lambda},  \label{eq:DvB}
\end{equation}
where $v_B$ is the butterfly velocity which characterizes the spatial growth of OTOCs \cite{Roberts:2014isa}:
\begin{equation}
\mathrm{tr}\left(\rho^{1/4} A(x,t) \rho^{1/4} B \rho^{1/4} A(x,t) \rho^{1/4} B\right) \sim 1 - \frac{1}{N} e^{\lambda (t-x/v_B)} + \cdots.
\end{equation}
The relationship (\ref{eq:DvB}) is profound: it links a readily measurable transport coefficient $D$ with two hard-to-measure coefficients characterizing chaos.  In holographic models, this relation is natural (especially for thermal diffusion) as $D$, $v_B$ and $\lambda$ are all calculable near the horizon.  But (\ref{eq:DvB}) has also been confirmed in kinetic theories \cite{Aleiner:2016eni, Grozdanov:2018atb} and other field-theoretic models \cite{Patel:2016wdy, Werman:2017abn, Patel:2017vfp}, so it may be more universal.  An outstanding challenge with linking (\ref{eq:DvB}) to Planckian resistivity bounds is that the natural \emph{bound} form of (\ref{eq:DvB}) is $D \lesssim v^2 \tau$ \cite{Lucas:2016yfl, Hartman:2017hhp},
with $v$ and $\tau$ characteristic many-body velocities and times that can be related to $v_B$ and $1/\lambda$ in some circumstances. This bound might seem to suggest that Planckian transport could be associated with minimum, not maximum, resistivity. However, in \cite{Delacretaz:2021ufg} an upper bound on the diffusivity in two dimensional CFTs was shown to lead to an upper bound on the thermalization rate that was even stronger than the Planckian bound (\ref{eq:planckianbound}).

There are a few possible routes to unification. The coming decade will hopefully see substantial progress on these questions, using a combination of insights and calculations from holographic and non-holographic models. The phenomenon of pole-skipping \cite{Grozdanov:2017ajz, Blake:2017ris, Blake:2018leo, Grozdanov:2020koi} in complex-wave number Green's functions has a clear link to chaos.   It has also been proposed that finite temperature operator size growth is the common link \cite{Lucas:2018wsc}.   We anticipate that holographic models may prove particularly fruitful in linking chaos to quasinormal mode decay rates, which are typically Planckian \cite{Horowitz:1999jd, Starinets:2002br, Motl:2003cd,Kovtun:2005ev, Festuccia:2008zx, Sybesma:2015oha}, perhaps most universally those associated with the stress tensor.  An important point is that while it is believed that Planckian decay times in holography are a consequence of near-horizon physics, a compelling and detail-independent understanding of this appears missing.

\section{Late time many-body quantum chaos}
\label{sec:spectral}

As we mentioned above, the statistical properties of energy eigenvalues provide one productive way of defining quantum chaos for an ensemble of Hamiltonians. This idea has roots in Wigner's ansatz for nuclear energy levels~\cite{10.2307/1970079} and was connected to classical chaos by the Bohigas-Giannoni-Schmit conjecture~\cite{PhysRevLett.52.1}. Given such an ensemble, e.g. a set of nuclei, we say that the ensemble is quantum chaotic if the energy eigenvalues look statistically like those of a Gaussian random matrix ensemble. The statistics can be characterized via the spectral form factor, defined as the ensemble average of $|\text{tr}(e^{-i H t/\hbar})|^2$. Chaotic systems are supposed to exhibit a sustained period of linear-in-time growth of the form factor which onsets after a non-universal early time regime.

This framework for quantum chaos is appealing in part because it identifies a non-trivial structure that appears in an enormous range of physical systems including few particle systems with chaotic classical dynamics, many-particle systems of interacting spins with no classical limit, strongly coupled quantum field theories, and even quantum models of black hole dynamics. These connections have inspired a fruitful exchange of ideas between high energy physics and many-body physics, including in the context of solvable models (e.g.~\cite{GG_2016,Bertini_2018,Cotler:2016fpe}) and effective theories (e.g.~\cite{Altland_2021}). There is also work relating random matrix universality to other phenomena associated with quantum chaos, such as hydrodynamics~\cite{Winer:2020gdp} and the statistics of energy eigenstates (see e.g.~\cite{DAlessio_2016}). Random matrix universality has also been used to study circuit complexity in chaotic systems~\cite{HJ_2017}, motivated by complexity growth conjectures in quantum gravity (e.g.~\cite{susskind2018lectures}).

As connections between different communities continue to blossom, many directions have emerged for further investigation. For example, classical chaos theory is deeply connected to transport physics (e.g.~\cite{dorfman_1999}), and it could be extremely useful if similar connections exist in chaotic quantum systems. This may relate to our discussion of bounds in the previous section. We can also expect significant progress on a number of other directions, the continued development of effective theories of quantum chaos, the establishment of equivalences between different manifestations of quantum chaos, the discovery of new connections between chaos and quantum sampling and optimization tasks, and new experiments probing quantum chaos using quantum computers.

\section{Quantum information of many-body systems}

In the past two decades, the ideas of quantum error-correction and tensor networks have played a key role in studies of static properties of many-body systems, such as the classification of quantum phases of matter and providing numerical ans\"atze for characterizing critical systems. Recent significant developments in quantum gravity have witnessed that the quantum error-correction interpretation~\cite{Almheiri:2015aa} and tensor network models~\cite{Pastawski15b, Hayden:2016aa} of the AdS/CFT correspondence can also revolutionize our understanding of quantum gravity, leading to proofs of the entanglement wedge reconstruction~\cite{Dong2016}, a sharper understanding of the black hole interior and a derivation of the Page curve~\cite{Almheiri19, Penington19}. Insights from tensor network toy models have also provided geometric interpretations of quantum circuit complexity growth in CFT wavefunctions~\cite{Hartman:2013qma, Stanford:2014aa, Roberts:2014isa}, and an improved understanding of complexity has shed new light on the black hole interior as well~\cite{Brown19}. Now that quantum gravity makes use of these ideas to study dynamical properties of quantum black holes, we naturally expect that these insights will give birth to new research avenues in the dynamics of quantum many-body systems. Indeed, tensor networks circuits, which are geometric interpretations of the wormholes in two-sided black holes, have found useful application as tractable toy models of dynamics arising in many-body quantum systems~\cite{Nahum:2017aa}. These developments have also stimulated revival of interests in pseudo-randomness and complexity growth from the quantum information community~\cite{Roberts:2017aa, Brandao19, Haferkamp21}.

Another successful development, fuelled by advancement of our understanding of quantum gravity, concerns the notion of quantum information scrambling~\cite{HP2007}. We have discussed the role of OTOCs in characterizing scrambling in sections \ref{sec:onset} and \ref{sec:bounds} above, where we also described how these ideas have gone far beyond their original context and found applications in various corners of studies of quantum many-body systems. From an information-theoretic perspective, these ideas have furthermore stimulated significant experimental efforts to observe the phenomena of quantum information scrambling by simulating information recovery from an old black hole \cite{Hosur:2015ylk} in laboratories with cold atoms~\cite{Yoshida:2017aa, Landsman:2019aa}.

%In the context of the black hole physics, scrambling refers to the phenomena where an effect of tiny perturbations gets amplified near the horizon due to the blueshift and alters the entanglement structure of the black hole in a drastic manner~\cite{Shenker:2013pqa}. In CFT Hamiltonians, scrambling can be diagnosed by out-of-time ordered correlation functions of the form $\langle V(0)W(t)V(0)W(t)\rangle$~\cite{Kit.KITP.2, Maldacena:2015waa} which can probe the growth of local operators under chaotic dynamics~\cite{Roberts:2014isa}.

%Furthermore, this has led to a proof of the quantum information recoverability from an old black hole by establishing an information theoretic relation between scrambling and recoverability~\cite{Hosur:2015ylk}.
%The concept of quantum information scrambling and out-of-time ordered correlation functions have gone far beyond their original context and found applications in various corners of studies of quantum many-body systems. 

Finally, let us mention one recent development which makes non-trivial use of modern quantum information theory and is likely to suggest a host further generalization on dynamics arising in many-body systems. The AdS/CFT correspondence predicts that any physical processes in the bulk can be holographically realized on the boundary quantum system. It is however far from obvious how bulk processes, such as scattering of two particles, are actually described on the boundary. In 2019, Alex May has discovered that the bulk local quantum information processing tasks can be holographically realized in boundary quantum systems only if there exist pre-shared entanglement of some special form in the boundary wavefunction~\cite{May:2019aa}. Namely, it has been found that local interactions in the bulk can be holographically realized only as non-local quantum computations~\cite{Vaidman:2003aa, Beigi:2011aa} where distributed entanglement needs to be cleverly used as resources. This “holographic task” paradigm applies modern quantum information concepts, such as the quantum cryptography, to the holographic interpretations of the bulk scattering processes and sharpens our understanding of the entanglement and the causal structure. It also enabled quantum gravity to make surprising predictions about quantum information theory~\cite{May:2020aa}. This new paradigm poses an important future question: how do the interactions on the bulk emerge dynamically from the preexisting entanglement in the boundary CFT?

\section{The SYK model and beyond}
\label{sec:SYK}

A successful strategy in theoretical physics for demonstrating and analyzing subtle effects with precision is to investigate solvable toy models in detail. We have already mentioned the SYK model in previous sections and here we elaborate on it in more detail. In 2015, Kitaev proposed a simple model of quantum holography \cite{Kit.KITP.2} based on an old spin model of Sachdev and Ye \cite{sachdev1993}, now known as the Sachdev-Ye-Kitaev (SYK) model. The SYK model is a quantum system with a large number of  Majorana fermions and random all-to-all interactions, displaying emergent conformal symmetry at low energy. Since then, inspiration from the SYK model has benefited both condensed matter and black hole physics. For the former, it advances the way we describe and understand the quantum phases without quasi-particles such as non-Fermi liquids \cite{chowdhury2021sachdev}. 
For example, high dimensional variants of the SYK model often show local criticality \cite{si2001locally,Faulkner:2009wj}, a characteristic tied to the holographic non-Fermi liquids \cite{Liu:2009dm, Faulkner:2011tm,  Iqbal:2011ae, hartnoll2018holographic}. 
For the latter, having a low energy sector identified with Jackiw-Teitelboin gravity \cite{J1984,T1983}, the SYK model boosts the recent progress in the connection between black holes and random matrices \cite{Cotler:2016fpe,Saad:2018bqo,Saad:2019lba}, which further contributes to the idea of replica wormholes \cite{Penington:2019kki,Almheiri:2019qdq}. 

A remarkable property of the SYK model essential in building the bridge between condensed matter and black holes is the maximal chaos: SYK is a rare example of quantum systems that verifiably saturate the chaos bound,  %\cite{Maldacena:2015waa} 
a phenomenon first observed in black holes %\cite{Shenker:2013pqa,Shenker:2014cwa} 
as we have described above. It is then tempting to speculate that such quantum systems share certain universal properties with gravity. Therefore, the SYK model and its variants are the perfect testing grounds for those new ideas inspired by black holes, among which a typical example is the relation between the transport and information scrambling \cite{Gu:2016oyy,Gu:2017ohj,Hartman:2017hhp,Blake:2017ris,Blake:2018leo}, discussed in section \ref{sec:bounds} above. 

Looking forward, we hope to extend the lessons we learned from the SYK model and black holes to more general many-body quantum systems. 
Regarding chaos, what is the general structure of OTOCs? Some progress has been made along this direction exploiting variants of the SYK model not necessarily saturating the chaos bound \cite{Gu:2018jsv,GKZ21}. On the other hand, for quantum systems without a large number of local degrees of freedom, such as spin chains \cite{lin2018out,xu2020accessing}, the theory of OTOCs is less developed. 
%Another related topic is the relation between OTOCs and more traditional diagnostics of quantum chaos such as level statistics: both of which have been related to gravity in specific contexts. However, the direct connection between the two is still mysterious. 
For holographers, the SYK model is neat but not perfect. It lacks a gap in the conformal dimension spectrum, losing the sub-AdS locality in its dual \cite{Heemskerk:2009pn,ZGK20}. 
How to construct the next generation of toy models that improve this aspect? 
Last but not least, the SYK model provides exciting chances to access black holes in table-top experiments \cite{franz2018mimicking}. Interesting proposals to realize traversable wormholes in the lab have been in progress \cite{Brown:2019hmk,Schuster:2021uvg,Blok:2020may}. There are certainly more opportunities to come in the next few years for vivid imaginations. 

\section{Properties and Dynamics of Quark-Gluon Plasma}
\label{sec:qgp}

Because an AdS black hole, whose horizon has a Hawking temperature $T_H$, is dual to the hot strongly coupled liquid phase, at temperature $T=T_H$, in a nonabelian gauge theory --- in the simplest case ${\cal N}=4$ SYM theory --- holographic methods have become powerful tools for gleaning qualitative, and in some cases semi-quantitative, insights into the quark-gluon
plasma (QGP) phase of QCD.  It is worth noting that hot QCD matter at temperatures above the crossover temperature $T_c$ at which hadrons formed 
is much more similar to  ${\cal N}=4$ SYM matter at nonzero temperature than the vacua of the two theories are; neither is confining, neither features chiral symmetry breaking, neither has well-defined quasi-particles.  This is one reason why holographic methods have made a particular impact on understanding QGP.
The second reason is that there were, and still are, many open zeroth-order questions about the properties of QGP, the dynamics via which it is formed in ultrarelativistic heavy ion collisions at the LHC and RHIC, and the dynamics associated with probing it with jets or heavy quarks or quarkonia and seeing how it responds to these probes.  Obtaining reliable from-first-principles results in a quantum field theory that is not QCD but that features a similar hot strongly coupled liquid phase has helped to advance our understanding of QGP itself, and of phenomena in heavy ion collisions. For reviews, see Refs.~\cite{Schafer:2009dj,Casalderrey-Solana:2011dxg,Adams:2012th,Jacak:2012dx,dwdw,Romatschke:2017ejr,Busza:2018rrf}.

The earliest success was the realization that in the 
strongly coupled ${\cal N}=4$ SYM plasma the pressure, energy density and entropy density ($s$) are all $3/4$ of what they would be at zero coupling~\cite{Gubser:1996de,Witten:1998zw}, which is not bad at all as a qualitative characterization of what we know from lattice QCD about the properties of QGP over a wide range of temperatures above $T_c$~\cite{Borsanyi:2013bia,HotQCD:2014kol}.  This by itself would have remained just a curiosity, since these thermodynamic quantities {\it can} be formulated as Euclidean lattice calculations that can be performed reliably in QCD itself.
Not long after, though, as we mentioned in section \ref{sec:hydro} above, came the calculation of the shear viscosity $\eta$, showing that it takes on the remarkably small value $\eta=s/(4\pi)$~\cite{Policastro:2001yc,Policastro:2002se,Kovtun:2004de} in 
a large class of gauge theory plasmas~\cite{Son:2002sd,Kovtun:2003wp}.
This was of more interest because transport properties cannot 
be formulated in terms of the Euclidean path integral which describes equilibrium thermodynamics; they describe the relaxation toward equilibrium.  Furthermore, the comparison between heavy ion collision measurements of how anisotropic droplets of QGP expand and hydrodynamic calculations of same has demonstrated that QGP in a range of temperatures above $T_c$ has a comparably small $\eta/s$~\cite{Teaney:2003kp,Romatschke:2007mq,Luzum:2008cw,Song:2010mg,Casalderrey-Solana:2011dxg,Heinz:2013th}, making it the most strongly coupled liquid known.  

From this starting point, the power of holography -- in particular the ability to use it to formulate far-from-equilibrium questions tractably in a strongly coupled quantum field theory -- has led to a continuing explosion of interest, and of results. Calculations of the drag force experienced by a heavy quark plowing through strongly coupled plasma as well as its diffusion~\cite{Herzog:2006gh,Gubser:2006bz,Casalderrey-Solana:2006fio,Gubser:2006nz,Casalderrey-Solana:2007ahi,Reiten:2019fta} and of the jet quenching parameter describing how a massless parton traversing the soup picks up transverse momentum~\cite{Liu:2006ug,DEramo:2010wup} 
and loses energy~\cite{Chesler:2007an,Chesler:2014jva,Chesler:2015nqz} have by now led to a hybrid model that combines insights derived from such holographic calculations with a perturbative QCD description of parton showers to describe many aspects of how jets interact with QGP, the wakes they leave behind in the QGP, the resolving power of the QGP,  and more and that has been confronted successfully with data from the LHC and RHIC on varied observables~\cite{Casalderrey-Solana:2014bpa,Casalderrey-Solana:2015vaa,Casalderrey-Solana:2016jvj,Hulcher:2017cpt,Casalderrey-Solana:2018wrw,Casalderrey-Solana:2019ubu,Casalderrey-Solana:2020rsj}. 
Early holographic calculations demonstrating rapid hydrodynamization, which is to say QGP formation, in collisions~\cite{Chesler:2010bi,Chesler:2013lia,Busza:2018rrf} have by now led to holographically derived benchmark calculations of the initial stages of
a heavy ion collision~\cite{Casalderrey-Solana:2013aba,vanderSchee:2013pia,Chesler:2016ceu,Grozdanov:2016zjj,Bantilan:2018vjv,Folkestad:2019lam}. Holography, and in particular the study of non-hydrodynamic modes, has also contributed quite considerably recently to the understanding of equilibration~\cite{Chesler:2008hg,Chesler:2009cy,Heller:2011ju,Keegan:2015avk,Heller:2016gbp,Grozdanov:2016vgg,Florkowski:2017olj,Casalderrey-Solana:2018rle,Grozdanov:2018gfx,Grozdanov:2019kge,Grozdanov:2019uhi,Berges:2020fwq,Grozdanov:2020koi,Heller:2020hnq,Grozdanov:2021gzh}. In all these cases and more, holographic calculational methods have become a part of the standard toolkit. For essentially any interesting phenomena in heavy ion collisions, if it has proved possible to ``bracket reality'' by framing one simplified calculation using perturbative QCD methods~\cite{Ghiglieri:2020dpq} and another via holography, better understanding has been the result. Results cited above provide many examples of the value of this approach; another is the demonstration via weakly coupled~\cite{Milhano:2015mng}, holographic~\cite{Rajagopal:2016uip,Brewer:2017fqy} and hybrid~\cite{Casalderrey-Solana:2016jvj,Casalderrey-Solana:2018wrw,Casalderrey-Solana:2019ubu,Brewer:2021hmh} 
calculations that, among jets with the same energy, those that are wider in angle and/or contain more subjets and/or fragment into more partons lose much more energy. One example yet to come is the calculation of the non-abelian electric field correlator in Ref.~\cite{Binder:2021otw} which is of central interest to computations of quarkonium transport in QGP~\cite{Yao:2021lus} and (!!) of the dark matter relic abundance in certain classes of theories of dark matter~\cite{Binder:2021otw}. These authors have computed the correlator at weak coupling; the complementary holographic computation for strongly coupled plasma is underway, is challenging, and will teach us a lot.

Dynamics at strong coupling, whether relaxation toward equilibrium or full-blown far-from-equilibrium dynamics, is not the only interesting domain that is, to date, beyond the reach of lattice QCD and hence fertile ground for holographic calculations. QCD at nonzero, and in particular at large, baryon chemical potential $\mu_B$ is another.  This domain is of considerable current interest from two perspectives.  The Beam Energy Scan at RHIC has just completed three years of data taking, with the goal of exploring the phase diagram at nonzero $\mu_B$ and $T$, looking for a critical point and the beginnings of a first order phase transition. And, the combination of LIGO/VIRGO measurements of neutron star binary inspirals and X-Ray measurements from NICER that are constraining neutron star radii is opening new windows onto the properties of dense nuclear matter and, possibly, cold dense quark matter~\cite{Annala:2017llu,Annala:2019puf,Annala:2021gom}.  In both these domains, interesting, open, zeroth-order questions abound.  And, we are now seeing pioneering attempts to frame holographic calculations with the aim of gaining qualitative, and maybe semi-quantitative, insights into strongly coupled first order phase transitions~\cite{Attems:2019yqn,Janik:2021jbq,Bea:2021ieq} and cold dense quark matter~\cite{Annala:2017tqz,Faedo:2018fjw,Henriksson:2019zph,Faedo:2019jlp,Hoyos:2020hmq,Jokela:2021vwy,Hoyos:2021uff}.

\section{Far-from-equilibrium dynamics} 
\label{sec:far}

Far-from-equilibrium problems come in a large number of varieties and are notoriously difficult to deal with, not to mention for quantum systems at strong couplings. Direct numerical simulations of quantum many-body systems are often limited to small sizes. Instead of  following the time evolution of a wave function (or density matrix)
of a system consisting of large numbers of constituents,  holographic duality convert such problems to evolutions in classical gravity  (see~\cite{Hubeny:2010ry,Liu:2018crr} for reviews) involving solving partial differential equations (PDEs), 
for which powerful techniques have been developed during the last two decades. 
Furthermore, {the} gravity description provides  new ways to organize physics of a system which could be particularly valuable in non-equilibrium contexts.  For example, the extra spatial dimension in the bulk, often referred to as the radial direction, can be considered as a geometrization of renormalization group flow of a boundary theory. By analyzing the gravity geometry at different  
radial locations, one can learn about the boundary system at different scales. 
There has been significant progress in treating many far-from-equilibrium problems using gravity techniques and much more can be expected in the future.

Deep connections between Einstein's equations and full nonlinear hydrodynamics were uncovered in~\cite{Bhattacharyya:2008jc}, and have yielded new understandings of hydrodynamics for systems with quantum anomalies~\cite{Erdmenger:2008rm,Banerjee:2008th,Son:2009tf}, which in turn found applications in the studies of the quark-gluon plasmas and various condensed matter systems. 
Thermalization from a variety of homogeneous and inhomogeneous far-from-equilibrium states can be treated on the gravity side as gravitational collapses to form black holes. Depending on the setup of the initial configurations for the collapse and observables one calculates, it is possible to gain insights into many different aspects of thermalization of an isolated quantum system. For example, it was found that local thermalization (i.e. non-conserved quantities become locally equilibrated) can happen extremely fast for strongly coupled systems, at time scales of order $1/T$ with $T$ the final equilibrium temperature~\cite{Chesler:2008hg,Chesler:2010bi,Heller:2012km}. This is the Planckian time discussed in section \ref{sec:bounds} above and provides an `explanation' for the extremely short time scales observed for the creation and thermalization of the quark-gluon plasma following heavy ion collisions 
at RHIC and LHC (see \cite{dwdw,Casalderrey-Solana:2011dxg} for reviews). It has also been found that hydrodynamics can become valid much earlier than previously expected, before a system isotropizes~\cite{Chesler:2009cy,Chesler:2010bi}. 
From the evolution of entanglement entropy of a spatial region, a new quantity called entanglement velocity has emerged as a valuable characterization of growth of entanglement during a thermalization process~\cite{Hartman:2013qma,Liu:2013iza}. There has also been support from gravity analysis~\cite{Mezei:2018jco} that dynamics of entanglement growth may allow a coarse-grained hydrodynamic description~\cite{Jonay:2018yei}.

Black holes have also been used to study turbulence in both ordinary fluids and superfluids, which in turn have led to new insights into black hole geometries. By constructing black hole solutions dual to turbulent flows, it has been argued that the event horizon of a  ``turbulent'' black hole is a fractal, with its fractal dimension determined by the Kolmogorov scaling law~\cite{Adams:2013vsa}. 
Turbulent flows of a $(2+1)$-dimensional superfluid have been constructed holographically as a chaotic gas of cosmic strings 
 in a hairy black hole geometry, which provides a valuable laboratory for studying many open questions in superfluid turbulence. For example, with the identification of a new energy dissipation mechanism through vortex dynamics, 
 it has been shown that turbulence of a strongly coupled $(2+1)$-dimensional superfluid can have the direct energy cascade~\cite{Adams:2012pj}. 

There remain many mysteries regarding how an isolated quantum many-body system thermalizes. For example, while it is widely believed that chaos should play a crucial role in the process, a precise picture is still lacking. 
Much is still to be understood regarding quantum informational properties of a system during thermalization process. 
On the one hand we expect that black holes will continue to provide a valuable laboratory for studying these questions. On the  
other hand progress in these questions will also yield new insights into black hole physics.

\section{Summary and further directions}

Before concluding, we should mention a few topics that have not been covered here. Some of these topics will be discussed in other white papers. Our focus has been on new ideas for many-body quantum systems that have emerged through the lens of black holes. We have not discussed the impact these ideas have had on our understanding of quantum gravity itself,  e.g.~\cite{Bousso:2022ntt}. We have also not discussed the vibrant interplay of the topics outlined here with more conventional methods in many-body physics, in particular state-of-the art numerics on quantum many-body systems such as \cite{PhysRevX.5.041041, PhysRevX.7.031059}. Finally, we have not covered several other points of connection between high energy and gravitational physics with many-body physics that have a more field theoretic flavor. These will be considered in separate white papers and include the development of large $N$ effective field theories for Fermi surfaces, the development of mathematical techniques to describe topological and `fracton' phases of matter, dualities in 2+1 dimensional systems and the extension of `bootstrap' methods to nonzero temperature.

To summarize our discussion, the interface between black hole physics and many-body quantum systems has developed significantly over the past decade. This interface is ultimately possible because of the intrinsically thermodynamic nature of classical gravity as well as the microscopic grounding of this thermodynamics in string-theoretic holographic duality. In this white paper we have outlined the ongoing development of sophisticated and interlinked many-body concepts built around notions from quantum information theory, quantum chaos and hydrodynamics. These concepts have led to new ways to probe fundamental gravitational physics. At the same time, the ability to perform controlled computations in a gravitational setting --- and in the closely related SYK model --- 
has illuminated the behavior of strongly quantum many-body systems more generally, motivating new calculations and experiments across several fields of physics including condensed matter, nuclear physics, ultracold atoms and quantum computation. Research at this interface is expected to remain vibrant over the coming decade.

Finally, looking further afield, there are several rather different points of connection between gravitation and string theory with many-body physics that remain to be fleshed out but may emerge as important new directions in the future. We will very briefly mention some of these directions.

One direction is a proper understanding of the string landscape \cite{Bousso:2000xa}. The rich structure of supersymmetric vacua and supersymmetric multi-centered black hole solutions in string theory calls out for a statistical physics perspective, possibly connecting to ideas developed to describe energy landscapes in glassy many-body systems \cite{Denef:2011ee}. It is plausible that a statistical understanding of the string landscape is necessary to address deep questions such as the microscopic nature of de Sitter entropy, cf.~\cite{Dong:2010pm}. More generally, given the major breakthroughs in the past decade relating black hole horizons to many-body physics, the mysteries of cosmological and in particular de Sitter horizons present themselves as the next frontier \cite{Anninos:2012qw}. There has been recent important progress understanding the statistical origin of the leading quantum corrections to the classical de Sitter entropy \cite{Anninos:2020hfj}.

In another direction, many-body systems that are dual to Einstein gravity involve large $N$ matrix degrees of freedom. These are an interesting class of quantum many-body systems in their own right and solving them will be essential to understand the emergence of semiclassical, gravitating spacetime. The time may be ripe to attack this challenge. Recent work has adapted machine-learning techniques from conventional many-body systems to find the ground state of a matrix quantum mechanics \cite{PhysRevX.10.011069}. Other works have successfully developed bootstrap methods for solving theories with matrix degrees of freedom \cite{Lin:2020mme, Han:2020bkb}. The quantum-mechanical bootstrap developed in \cite{Han:2020bkb} may prove suitable for bootstrapping more conventional lattice many-body systems, with promising initial results reported in \cite{han2020quantum}. This illustrates the possibility that getting to grips with matrix quantum mechanics may well generate innovative ideas with wide-ranging impact, analogously to those we have discussed throughout this paper.

Finally, the more refined quantum chaos and quantum information-theoretic probes of many-body systems can sometimes be most simply phrased in terms of the interior rather than the exterior of the dual black hole. However, the relationship of the rich classical interior dynamics of black holes to the many-body system is fundamentally not understood. Recent works have begun to approach this question by identifying robust features of holographic black hole interiors, going beyond the conventionally used Schwarzschild and Reissner-Nordstrom solutions \cite{Frenkel:2020ysx, Hartnoll:2020rwq, Hartnoll:2020fhc}. Furthermore, the emergence of interior time has very recently been shown to be closely tied up with the $N \to \infty$ limit of the dual many-body system \cite{Leutheusser:2021qhd, Leutheusser:2021frk}. These works may lay the foundation for future maps between interior gravitational evolution and quantum many-body dynamics.

\section*{Acknowledgements}

YG is supported by the Simons Foundation through the “It from Qubit” program. SAH is supported by Simons Investigator award \#620869 and STFC consolidated grant ST/T000694/1.  HL is supported by the Office of High Energy Physics of U.S. Department of Energy under grant Contract Number  DE-SC0012567 and DE-SC0020360 (MIT contract \# 578218).
AL is supported by a Research Fellowship from the Alfred P. Sloan Foundation, by AFOSR Grant FA9550-21-1-0195, by NSF CAREER Grant DMR-2145544, and by the Gordon and Betty Moore Foundation's EPiQS Initiative under Grant GBMF10279. 
The work of KR is supported in part by the U.S.~Department of Energy, Office of Science, Office of Nuclear Physics grant DE-SC0011090. The work of BS is supported in part by the AFOSR under grant  FA9550-19-1-0360. BY is supported in part by Perimeter Institute for Theoretical Physics. Research at Perimeter Institute is supported in part by the Government of Canada through the Department of Innovation, Science and Economic Development Canada and by the Province of Ontario through the Ministry of Colleges and Universities.

\bibliographystyle{ourbst}
\bibliography{snowmass_bib}

\providecommand{\href}[2]{#2}\begingroup\raggedright\begin{thebibliography}{100}

\bibitem{Policastro:2001yc}
G.~Policastro, D.~T. Son and A.~O. Starinets, {{The Shear viscosity of strongly
  coupled N=4 supersymmetric Yang-Mills plasma}},
  \href{http://dx.doi.org/10.1103/PhysRevLett.87.081601}{Phys. Rev. Lett. {\bf
  87}, 081601, 2001},
  [\href{http://arxiv.org/abs/arXiv:hep-th/0104066}{{arXiv:hep-th/0104066}}].

\bibitem{Kovtun:2003wp}
P.~Kovtun, D.~T. Son and A.~O. Starinets, {{Holography and hydrodynamics:
  Diffusion on stretched horizons}},
  \href{http://dx.doi.org/10.1088/1126-6708/2003/10/064}{JHEP {\bf 0310}, 064,
  2003},
  [\href{http://arxiv.org/abs/arXiv:hep-th/0309213}{{arXiv:hep-th/0309213
  [hep-th]}}].

\bibitem{Kovtun:2004de}
P.~Kovtun, D.~T. Son and A.~O. Starinets, {{Viscosity in strongly interacting
  quantum field theories from black hole physics}},
  \href{http://dx.doi.org/10.1103/PhysRevLett.94.111601}{Phys.Rev.Lett. {\bf
  94}, 111601, 2005},
  [\href{http://arxiv.org/abs/arXiv:hep-th/0405231}{{arXiv:hep-th/0405231
  [hep-th]}}].

\bibitem{Shuryak:2008eq}
E.~Shuryak, {{Physics of Strongly coupled Quark-Gluon Plasma}},
  \href{http://dx.doi.org/10.1016/j.ppnp.2008.09.001}{Prog. Part. Nucl. Phys.
  {\bf 62}, 48--101, 2009},
  [\href{http://arxiv.org/abs/arXiv:0807.3033}{{arXiv:0807.3033 [hep-ph]}}].

\bibitem{Cao_2010}
C.~Cao, E.~Elliott, J.~Joseph, H.~Wu, J.~Petricka, T.~Sch\"afer and J.~E.
  Thomas, {{Universal Quantum Viscosity in a Unitary Fermi Gas}},
  \href{http://dx.doi.org/10.1126/science.1195219}{Science {\bf 331}, 58,
  2011}, [\href{http://arxiv.org/abs/arXiv:1007.2625}{{arXiv:1007.2625
  [cond-mat.quant-gas]}}].

\bibitem{muller2009}
M.~{M\"uller}, J.~Schmalian and L.~Fritz, {{Graphene - a nearly perfect
  fluid}}, {Phys. Rev. Lett. {\bf 103}, 025301, 2009},
  [\href{http://arxiv.org/abs/arXiv:0903.4178}{{arXiv:0903.4178
  [cond-mat.str-el]}}].

\bibitem{Crossno1058}
J.~Crossno, J.~K. Shi, K.~Wang, X.~Liu, A.~Harzheim, A.~Lucas, S.~Sachdev,
  P.~Kim, T.~Taniguchi, K.~Watanabe, T.~A. Ohki and K.~C. Fong, {{Observation
  of the Dirac fluid and the breakdown of the Wiedemann-Franz law in
  graphene}}, \href{http://dx.doi.org/10.1126/science.aad0343}{Science {\bf
  351}, 1058--1061, 2016},
  [\href{http://arxiv.org/abs/arXiv:1509.04713}{{arXiv:1509.04713}}].

\bibitem{Ku:2019lgj}
M.~J.~H. Ku et~al., {{Imaging viscous flow of the Dirac fluid in graphene}},
  \href{http://dx.doi.org/10.1038/s41586-020-2507-2}{Nature {\bf 583},
  537--541, 2020},
  [\href{http://arxiv.org/abs/arXiv:1905.10791}{{arXiv:1905.10791
  [cond-mat.mes-hall]}}].

\bibitem{Bandurin1055}
D.~A. Bandurin, I.~Torre, R.~K. Kumar, M.~Ben~Shalom, A.~Tomadin, A.~Principi,
  G.~H. Auton, E.~Khestanova, K.~S. Novoselov, I.~V. Grigorieva, L.~A.
  Ponomarenko, A.~K. Geim and M.~Polini, {{Negative local resistance caused by
  viscous electron backflow in graphene}},
  \href{http://dx.doi.org/10.1126/science.aad0201}{Science {\bf 351},
  1055--1058, 2016},
  [\href{http://arxiv.org/abs/arXiv:1509.04165}{{arXiv:1509.04165}}].

\bibitem{Krishna_Kumar_2017}
R.~K. Kumar et~al., {Superballistic flow of viscous electron fluid through
  graphene constrictions}, \href{http://dx.doi.org/10.1038/nphys4240}{Nature
  Physics {\bf 13}, 1182, 2017},
  [\href{http://arxiv.org/abs/arXiv:1703.06672}{{arXiv:1703.06672
  [cond-mat.mes-hall]}}].

\bibitem{sulpizio}
J.~A. {Sulpizio}, L.~{Ella}, A.~{Rozen}, J.~{Birkbeck}, D.~J. {Perello},
  D.~{Dutta}, M.~{Ben-Shalom}, T.~{Taniguchi}, K.~{Watanabe}, T.~{Holder},
  R.~{Queiroz}, A.~{Principi}, A.~{Stern} et~al., {{Visualizing Poiseuille flow
  of hydrodynamic electrons}},
  \href{http://dx.doi.org/10.1038/s41586-019-1788-9}{Nature {\bf 576}, 75--79,
  2019}, [\href{http://arxiv.org/abs/arXiv:1905.11662}{{arXiv:1905.11662
  [cond-mat.mes-hall]}}].

\bibitem{Lucas:2017idv}
A.~Lucas and K.~C. Fong, {{Hydrodynamics of electrons in graphene}},
  \href{http://dx.doi.org/10.1088/1361-648X/aaa274}{J. Phys. Condens. Matter
  {\bf 30}, 053001, 2018},
  [\href{http://arxiv.org/abs/arXiv:1710.08425}{{arXiv:1710.08425
  [cond-mat.str-el]}}].

\bibitem{Gubser:2008px}
S.~S. Gubser, {{Breaking an Abelian gauge symmetry near a black hole horizon}},
  \href{http://dx.doi.org/10.1103/PhysRevD.78.065034}{Phys. Rev. D {\bf 78},
  065034, 2008}, [\href{http://arxiv.org/abs/arXiv:0801.2977}{{arXiv:0801.2977
  [hep-th]}}].

\bibitem{Hartnoll:2008vx}
S.~A. Hartnoll, C.~P. Herzog and G.~T. Horowitz, {{Building a Holographic
  Superconductor}},
  \href{http://dx.doi.org/10.1103/PhysRevLett.101.031601}{Phys. Rev. Lett. {\bf
  101}, 031601, 2008},
  [\href{http://arxiv.org/abs/arXiv:0803.3295}{{arXiv:0803.3295 [hep-th]}}].

\bibitem{Hartnoll:2008kx}
S.~A. Hartnoll, C.~P. Herzog and G.~T. Horowitz, {{Holographic
  Superconductors}},
  \href{http://dx.doi.org/10.1088/1126-6708/2008/12/015}{JHEP {\bf 12}, 015,
  2008}, [\href{http://arxiv.org/abs/arXiv:0810.1563}{{arXiv:0810.1563
  [hep-th]}}].

\bibitem{Nakamura:2009tf}
S.~Nakamura, H.~Ooguri and C.-S. Park, {{Gravity Dual of Spatially Modulated
  Phase}}, \href{http://dx.doi.org/10.1103/PhysRevD.81.044018}{Phys. Rev. D
  {\bf 81}, 044018, 2010},
  [\href{http://arxiv.org/abs/arXiv:0911.0679}{{arXiv:0911.0679 [hep-th]}}].

\bibitem{Donos:2011bh}
A.~Donos and J.~P. Gauntlett, {{Holographic striped phases}},
  \href{http://dx.doi.org/10.1007/JHEP08(2011)140}{JHEP {\bf 08}, 140, 2011},
  [\href{http://arxiv.org/abs/arXiv:1106.2004}{{arXiv:1106.2004 [hep-th]}}].

\bibitem{Denef:2009tp}
F.~Denef and S.~A. Hartnoll, {{Landscape of superconducting membranes}},
  \href{http://dx.doi.org/10.1103/PhysRevD.79.126008}{Phys. Rev. D {\bf 79},
  126008, 2009}, [\href{http://arxiv.org/abs/arXiv:0901.1160}{{arXiv:0901.1160
  [hep-th]}}].

\bibitem{Liu:2009dm}
H.~Liu, J.~McGreevy and D.~Vegh, {{Non-Fermi liquids from holography}},
  \href{http://dx.doi.org/10.1103/PhysRevD.83.065029}{Phys. Rev. D {\bf 83},
  065029, 2011}, [\href{http://arxiv.org/abs/arXiv:0903.2477}{{arXiv:0903.2477
  [hep-th]}}].

\bibitem{Cubrovic:2009ye}
M.~Cubrovic, J.~Zaanen and K.~Schalm, {{String Theory, Quantum Phase
  Transitions and the Emergent Fermi-Liquid}},
  \href{http://dx.doi.org/10.1126/science.1174962}{Science {\bf 325}, 439--444,
  2009}, [\href{http://arxiv.org/abs/arXiv:0904.1993}{{arXiv:0904.1993
  [hep-th]}}].

\bibitem{Faulkner:2009wj}
T.~Faulkner, H.~Liu, J.~McGreevy and D.~Vegh, {{Emergent quantum criticality,
  Fermi surfaces, and AdS(2)}},
  \href{http://dx.doi.org/10.1103/PhysRevD.83.125002}{Phys. Rev. D {\bf 83},
  125002, 2011}, [\href{http://arxiv.org/abs/arXiv:0907.2694}{{arXiv:0907.2694
  [hep-th]}}].

\bibitem{faulkner2010strange}
T.~Faulkner, N.~Iqbal, H.~Liu, J.~McGreevy and D.~Vegh, {Strange metal
  transport realized by gauge/gravity duality}, {Science {\bf 329}, 1043--1047,
  2010}.

\bibitem{Iqbal:2011in}
N.~Iqbal, H.~Liu and M.~Mezei, {{Semi-local quantum liquids}},
  \href{http://dx.doi.org/10.1007/JHEP04(2012)086}{JHEP {\bf 04}, 086, 2012},
  [\href{http://arxiv.org/abs/arXiv:1105.4621}{{arXiv:1105.4621 [hep-th]}}].

\bibitem{Hartnoll:2012rj}
S.~A. Hartnoll and D.~M. Hofman, {{Locally Critical Resistivities from Umklapp
  Scattering}}, \href{http://dx.doi.org/10.1103/PhysRevLett.108.241601}{Phys.
  Rev. Lett. {\bf 108}, 241601, 2012},
  [\href{http://arxiv.org/abs/arXiv:1201.3917}{{arXiv:1201.3917 [hep-th]}}].

\bibitem{Kachru:2008yh}
S.~Kachru, X.~Liu and M.~Mulligan, {{Gravity duals of Lifshitz-like fixed
  points}}, \href{http://dx.doi.org/10.1103/PhysRevD.78.106005}{Phys. Rev. D
  {\bf 78}, 106005, 2008},
  [\href{http://arxiv.org/abs/arXiv:0808.1725}{{arXiv:0808.1725 [hep-th]}}].

\bibitem{Son:2008ye}
D.~T. Son, {{Toward an AdS/cold atoms correspondence: A Geometric realization
  of the Schrodinger symmetry}},
  \href{http://dx.doi.org/10.1103/PhysRevD.78.046003}{Phys. Rev. D {\bf 78},
  046003, 2008}, [\href{http://arxiv.org/abs/arXiv:0804.3972}{{arXiv:0804.3972
  [hep-th]}}].

\bibitem{Balasubramanian:2008dm}
K.~Balasubramanian and J.~McGreevy, {{Gravity duals for non-relativistic
  CFTs}}, \href{http://dx.doi.org/10.1103/PhysRevLett.101.061601}{Phys. Rev.
  Lett. {\bf 101}, 061601, 2008},
  [\href{http://arxiv.org/abs/arXiv:0804.4053}{{arXiv:0804.4053 [hep-th]}}].

\bibitem{Charmousis:2010zz}
C.~Charmousis, B.~Gouteraux, B.~S. Kim, E.~Kiritsis and R.~Meyer, {{Effective
  Holographic Theories for low-temperature condensed matter systems}},
  \href{http://dx.doi.org/10.1007/JHEP11(2010)151}{JHEP {\bf 11}, 151, 2010},
  [\href{http://arxiv.org/abs/arXiv:1005.4690}{{arXiv:1005.4690 [hep-th]}}].

\bibitem{Gouteraux:2011ce}
B.~Gouteraux and E.~Kiritsis, {{Generalized Holographic Quantum Criticality at
  Finite Density}}, \href{http://dx.doi.org/10.1007/JHEP12(2011)036}{JHEP {\bf
  12}, 036, 2011},
  [\href{http://arxiv.org/abs/arXiv:1107.2116}{{arXiv:1107.2116 [hep-th]}}].

\bibitem{Huijse:2011ef}
L.~Huijse, S.~Sachdev and B.~Swingle, {{Hidden Fermi surfaces in compressible
  states of gauge-gravity duality}},
  \href{http://dx.doi.org/10.1103/PhysRevB.85.035121}{Phys. Rev. B {\bf 85},
  035121, 2012}, [\href{http://arxiv.org/abs/arXiv:1112.0573}{{arXiv:1112.0573
  [cond-mat.str-el]}}].

\bibitem{Iqbal:2011ae}
N.~Iqbal, H.~Liu and M.~Mezei, {{Lectures on holographic non-Fermi liquids and
  quantum phase transitions}},  in \emph{{Theoretical Advanced Study Institute
  in Elementary Particle Physics}: {String theory and its Applications: From
  meV to the Planck Scale}}, pp.~707--816, 2011.
\newblock [\href{http://arxiv.org/abs/arXiv:1110.3814}{{arXiv:1110.3814
  [hep-th]}}].

\bibitem{Casalderrey-Solana:2011dxg}
J.~Casalderrey-Solana, H.~Liu, D.~Mateos, K.~Rajagopal and U.~A. Wiedemann,
  \emph{{Gauge/String Duality, Hot QCD and Heavy Ion Collisions}}.
\newblock Cambridge University Press, 2014,
  \href{http://dx.doi.org/10.1017/CBO9781139136747}{10.1017/CBO9781139136747}.

\bibitem{Zaanen:2015oix}
J.~Zaanen, Y.-W. Sun, Y.~Liu and K.~Schalm, \emph{{Holographic Duality in
  Condensed Matter Physics}}.
\newblock Cambridge Univ. Press, 2015.

\bibitem{hartnoll2018holographic}
S.~A. Hartnoll, A.~Lucas and S.~Sachdev, \emph{Holographic quantum matter}.
\newblock MIT press, 2018.

\bibitem{Liu:2020rrn}
H.~Liu and J.~Sonner, {{Quantum many-body physics from a gravitational lens}},
  \href{http://dx.doi.org/10.1038/s42254-020-0225-1}{Nature Rev. Phys. {\bf 2},
  615--633, 2020},
  [\href{http://arxiv.org/abs/arXiv:2004.06159}{{arXiv:2004.06159 [hep-th]}}].

\bibitem{Zaanen:2021llz}
J.~Zaanen, {{Lectures on quantum supreme matter}},
  [\href{http://arxiv.org/abs/arXiv:2110.00961}{{arXiv:2110.00961
  [cond-mat.str-el]}}].

\bibitem{Hartnoll:2014cua}
S.~A. Hartnoll and J.~E. Santos, {{Disordered horizons: Holography of randomly
  disordered fixed points}},
  \href{http://dx.doi.org/10.1103/PhysRevLett.112.231601}{Phys. Rev. Lett. {\bf
  112}, 231601, 2014},
  [\href{http://arxiv.org/abs/arXiv:1402.0872}{{arXiv:1402.0872 [hep-th]}}].

\bibitem{Narovlansky:2018muj}
V.~Narovlansky and O.~Aharony, {{Renormalization Group in Field Theories with
  Quantum Quenched Disorder}},
  \href{http://dx.doi.org/10.1103/PhysRevLett.121.071601}{Phys. Rev. Lett. {\bf
  121}, 071601, 2018},
  [\href{http://arxiv.org/abs/arXiv:1803.08529}{{arXiv:1803.08529
  [cond-mat.str-el]}}].

\bibitem{Ganesan:2021gun}
K.~Ganesan, A.~Lucas and L.~Radzihovsky, {{Renormalization group in quantum
  critical theories with Harris-marginal disorder}},
  [\href{http://arxiv.org/abs/arXiv:2110.11978}{{arXiv:2110.11978 [hep-th]}}].

\bibitem{Amoretti:2018tzw}
A.~Amoretti, D.~Are\'an, B.~Gout\'eraux and D.~Musso, {{Universal relaxation in
  a holographic metallic density wave phase}},
  \href{http://dx.doi.org/10.1103/PhysRevLett.123.211602}{Phys. Rev. Lett. {\bf
  123}, 211602, 2019},
  [\href{http://arxiv.org/abs/arXiv:1812.08118}{{arXiv:1812.08118 [hep-th]}}].

\bibitem{Delacretaz:2021qqu}
L.~V. Delacr\'etaz, B.~Gout\'eraux and V.~Ziogas, {{Damping of Pseudo-Goldstone
  Fields}},  [\href{http://arxiv.org/abs/arXiv:2111.13459}{{arXiv:2111.13459
  [hep-th]}}].

\bibitem{Armas:2021vku}
J.~Armas, A.~Jain and R.~Lier, {{Approximate symmetries, pseudo-Goldstones, and
  the second law of thermodynamics}},
  [\href{http://arxiv.org/abs/arXiv:2112.14373}{{arXiv:2112.14373 [hep-th]}}].

\bibitem{Andrade:2017ghg}
T.~Andrade, A.~Krikun, K.~Schalm and J.~Zaanen, {{Doping the Holographic Mott
  Insulator}}, \href{http://dx.doi.org/10.1038/s41567-018-0217-6}{Nature Phys.
  {\bf 14}, 1049--1055, 2018},
  [\href{http://arxiv.org/abs/arXiv:1710.05791}{{arXiv:1710.05791 [hep-th]}}].

\bibitem{Sachdev:2010um}
S.~Sachdev, {{Holographic metals and the fractionalized Fermi liquid}},
  \href{http://dx.doi.org/10.1103/PhysRevLett.105.151602}{Phys. Rev. Lett. {\bf
  105}, 151602, 2010},
  [\href{http://arxiv.org/abs/arXiv:1006.3794}{{arXiv:1006.3794 [hep-th]}}].

\bibitem{1969Larkin}
A.~I. {Larkin} and Y.~N. {Ovchinnikov}, {{Quasiclassical Method in the Theory
  of Superconductivity}}, {Soviet Journal of Experimental and Theoretical
  Physics {\bf 28}, 1200, 1969}.

\bibitem{Shenker:2013pqa}
S.~H. Shenker and D.~Stanford, {{Black holes and the butterfly effect}},
  \href{http://dx.doi.org/10.1007/JHEP03(2014)067}{JHEP {\bf 03}, 067, 2014},
  [\href{http://arxiv.org/abs/arXiv:1306.0622}{{arXiv:1306.0622 [hep-th]}}].

\bibitem{Roberts:2014isa}
D.~A. Roberts, D.~Stanford and L.~Susskind, {{Localized shocks}},
  \href{http://dx.doi.org/10.1007/JHEP03(2015)051}{JHEP {\bf 03}, 051, 2015},
  [\href{http://arxiv.org/abs/arXiv:1409.8180}{{arXiv:1409.8180 [hep-th]}}].

\bibitem{Shenker:2014cwa}
S.~H. Shenker and D.~Stanford, {{Stringy effects in scrambling}},
  \href{http://dx.doi.org/10.1007/JHEP05(2015)132}{JHEP {\bf 05}, 132, 2015},
  [\href{http://arxiv.org/abs/arXiv:1412.6087}{{arXiv:1412.6087 [hep-th]}}].

\bibitem{Kit.KITP.2}
A.~Kitaev, ``A simple model of quantum holography.'' Talks at KITP
  \url{http://online.kitp.ucsb.edu/online/entangled15/kitaev/} and
  \url{http://online.kitp.ucsb.edu/online/entangled15/kitaev2/}, 2015.

\bibitem{sachdev1993}
S.~Sachdev and J.~Ye, {Gapless spin-fluid ground state in a random quantum
  heisenberg magnet},
  \href{http://dx.doi.org/10.1103/physrevlett.70.3339}{Physical Review Letters
  {\bf 70}, 3339–3342, 1993},
  [\href{http://arxiv.org/abs/arXiv:cond-mat/9212030}{{arXiv:cond-mat/9212030
  [cond-mat]}}].

\bibitem{Kitaev:2017awl}
A.~Kitaev and S.~J. Suh, {{The soft mode in the Sachdev-Ye-Kitaev model and its
  gravity dual}}, \href{http://dx.doi.org/10.1007/JHEP05(2018)183}{JHEP {\bf
  05}, 183, 2018},
  [\href{http://arxiv.org/abs/arXiv:1711.08467}{{arXiv:1711.08467 [hep-th]}}].

\bibitem{Maldacena:2016hyu}
J.~Maldacena and D.~Stanford, {{Remarks on the Sachdev-Ye-Kitaev model}},
  \href{http://dx.doi.org/10.1103/PhysRevD.94.106002}{Phys. Rev. D {\bf 94},
  106002, 2016},
  [\href{http://arxiv.org/abs/arXiv:1604.07818}{{arXiv:1604.07818 [hep-th]}}].

\bibitem{Nahum:2017yvy}
A.~Nahum, S.~Vijay and J.~Haah, {{Operator Spreading in Random Unitary
  Circuits}}, \href{http://dx.doi.org/10.1103/PhysRevX.8.021014}{Phys. Rev. X
  {\bf 8}, 021014, 2018},
  [\href{http://arxiv.org/abs/arXiv:1705.08975}{{arXiv:1705.08975
  [cond-mat.str-el]}}].

\bibitem{vonKeyserlingk:2017dyr}
C.~von Keyserlingk, T.~Rakovszky, F.~Pollmann and S.~Sondhi, {{Operator
  hydrodynamics, OTOCs, and entanglement growth in systems without conservation
  laws}}, \href{http://dx.doi.org/10.1103/PhysRevX.8.021013}{Phys. Rev. X {\bf
  8}, 021013, 2018},
  [\href{http://arxiv.org/abs/arXiv:1705.08910}{{arXiv:1705.08910
  [cond-mat.str-el]}}].

\bibitem{Khemani:2017nda}
V.~Khemani, A.~Vishwanath and D.~A. Huse, {{Operator spreading and the
  emergence of dissipation in unitary dynamics with conservation laws}},
  \href{http://dx.doi.org/10.1103/PhysRevX.8.031057}{Phys. Rev. X {\bf 8},
  031057, 2018},
  [\href{http://arxiv.org/abs/arXiv:1710.09835}{{arXiv:1710.09835
  [cond-mat.stat-mech]}}].

\bibitem{Swingle:2016jdj}
B.~Swingle and D.~Chowdhury, {{Slow scrambling in disordered quantum systems}},
  \href{http://dx.doi.org/10.1103/PhysRevB.95.060201}{Phys. Rev. B {\bf 95},
  060201, 2017},
  [\href{http://arxiv.org/abs/arXiv:1608.03280}{{arXiv:1608.03280
  [cond-mat.str-el]}}].

\bibitem{Turiaci:2016cvo}
G.~Turiaci and H.~Verlinde, {{On CFT and Quantum Chaos}},
  \href{http://dx.doi.org/10.1007/JHEP12(2016)110}{JHEP {\bf 12}, 110, 2016},
  [\href{http://arxiv.org/abs/arXiv:1603.03020}{{arXiv:1603.03020 [hep-th]}}].

\bibitem{Gu:2016oyy}
Y.~Gu, X.-L. Qi and D.~Stanford, {{Local criticality, diffusion and chaos in
  generalized Sachdev-Ye-Kitaev models}},
  \href{http://dx.doi.org/10.1007/JHEP05(2017)125}{JHEP {\bf 05}, 125, 2017},
  [\href{http://arxiv.org/abs/arXiv:1609.07832}{{arXiv:1609.07832 [hep-th]}}].

\bibitem{Maldacena:2015waa}
J.~Maldacena, S.~H. Shenker and D.~Stanford, {{A bound on chaos}},
  \href{http://dx.doi.org/10.1007/JHEP08(2016)106}{JHEP {\bf 08}, 106, 2016},
  [\href{http://arxiv.org/abs/arXiv:1503.01409}{{arXiv:1503.01409 [hep-th]}}].

\bibitem{Blake:2016wvh}
M.~Blake, {{Universal Charge Diffusion and the Butterfly Effect}},
  \href{http://dx.doi.org/10.1103/PhysRevLett.117.091601}{Phys. Rev. Lett. {\bf
  117}, 091601, 2016},
  [\href{http://arxiv.org/abs/arXiv:1603.08510}{{arXiv:1603.08510 [hep-th]}}].

\bibitem{Blake:2016sud}
M.~Blake, {{Universal Diffusion in Incoherent Black Holes}},
  [\href{http://arxiv.org/abs/arXiv:1604.01754}{{arXiv:1604.01754 [hep-th]}}].

\bibitem{Grozdanov:2017ajz}
S.~Grozdanov, K.~Schalm and V.~Scopelliti, {{Black hole scrambling from
  hydrodynamics}},
  \href{http://dx.doi.org/10.1103/PhysRevLett.120.231601}{Phys. Rev. Lett. {\bf
  120}, 231601, 2018},
  [\href{http://arxiv.org/abs/arXiv:1710.00921}{{arXiv:1710.00921 [hep-th]}}].

\bibitem{Blake:2017ris}
M.~Blake, H.~Lee and H.~Liu, {{A quantum hydrodynamical description for
  scrambling and many-body chaos}},
  \href{http://dx.doi.org/10.1007/JHEP10(2018)127}{JHEP {\bf 10}, 127, 2018},
  [\href{http://arxiv.org/abs/arXiv:1801.00010}{{arXiv:1801.00010 [hep-th]}}].

\bibitem{Blake:2018leo}
M.~Blake, R.~A. Davison, S.~Grozdanov and H.~Liu, {{Many-body chaos and energy
  dynamics in holography}},
  \href{http://dx.doi.org/10.1007/JHEP10(2018)035}{JHEP {\bf 10}, 035, 2018},
  [\href{http://arxiv.org/abs/arXiv:1809.01169}{{arXiv:1809.01169 [hep-th]}}].

\bibitem{Haehl:2018izb}
F.~M. Haehl and M.~Rozali, {{Effective Field Theory for Chaotic CFTs}},
  \href{http://dx.doi.org/10.1007/JHEP10(2018)118}{JHEP {\bf 10}, 118, 2018},
  [\href{http://arxiv.org/abs/arXiv:1808.02898}{{arXiv:1808.02898 [hep-th]}}].

\bibitem{Haehl:2019eae}
F.~M. Haehl, W.~Reeves and M.~Rozali, {{Reparametrization modes, shadow
  operators, and quantum chaos in higher-dimensional CFTs}},
  \href{http://dx.doi.org/10.1007/JHEP11(2019)102}{JHEP {\bf 11}, 102, 2019},
  [\href{http://arxiv.org/abs/arXiv:1909.05847}{{arXiv:1909.05847 [hep-th]}}].

\bibitem{Blake:2021wqj}
M.~Blake and H.~Liu, {{On systems of maximal quantum chaos}},
  \href{http://dx.doi.org/10.1007/JHEP05(2021)229}{JHEP {\bf 05}, 229, 2021},
  [\href{http://arxiv.org/abs/arXiv:2102.11294}{{arXiv:2102.11294 [hep-th]}}].

\bibitem{Gu:2018jsv}
Y.~Gu and A.~Kitaev, {{On the relation between the magnitude and exponent of
  OTOCs}}, \href{http://dx.doi.org/10.1007/JHEP02(2019)075}{JHEP {\bf 02}, 075,
  2019}, [\href{http://arxiv.org/abs/arXiv:1812.00120}{{arXiv:1812.00120
  [hep-th]}}].

\bibitem{Cotler:2016fpe}
J.~S. Cotler, G.~Gur-Ari, M.~Hanada, J.~Polchinski, P.~Saad, S.~H. Shenker,
  D.~Stanford, A.~Streicher and M.~Tezuka, {{Black Holes and Random Matrices}},
  \href{http://dx.doi.org/10.1007/JHEP05(2017)118}{JHEP {\bf 05}, 118, 2017},
  [\href{http://arxiv.org/abs/arXiv:1611.04650}{{arXiv:1611.04650 [hep-th]}}].

\bibitem{Saad:2018bqo}
P.~Saad, S.~H. Shenker and D.~Stanford, {{A semiclassical ramp in SYK and in
  gravity}},  [\href{http://arxiv.org/abs/arXiv:1806.06840}{{arXiv:1806.06840
  [hep-th]}}].

\bibitem{Winer:2020gdp}
M.~Winer and B.~Swingle, {{Hydrodynamic Theory of the Connected Spectral Form
  Factor}},  [\href{http://arxiv.org/abs/arXiv:2012.01436}{{arXiv:2012.01436
  [cond-mat.stat-mech]}}].

\bibitem{Zhou:2018myl}
T.~Zhou and A.~Nahum, {{Emergent statistical mechanics of entanglement in
  random unitary circuits}},
  \href{http://dx.doi.org/10.1103/PhysRevB.99.174205}{Phys. Rev. B {\bf 99},
  174205, 2019},
  [\href{http://arxiv.org/abs/arXiv:1804.09737}{{arXiv:1804.09737
  [cond-mat.stat-mech]}}].

\bibitem{Jonay:2018yei}
C.~Jonay, D.~A. Huse and A.~Nahum, {{Coarse-grained dynamics of operator and
  state entanglement}},
  [\href{http://arxiv.org/abs/arXiv:1803.00089}{{arXiv:1803.00089
  [cond-mat.stat-mech]}}].

\bibitem{Zhou:2019pob}
T.~Zhou and A.~Nahum, {{Entanglement Membrane in Chaotic Many-Body Systems}},
  \href{http://dx.doi.org/10.1103/PhysRevX.10.031066}{Phys. Rev. X {\bf 10},
  031066, 2020},
  [\href{http://arxiv.org/abs/arXiv:1912.12311}{{arXiv:1912.12311
  [cond-mat.str-el]}}].

\bibitem{Mezei:2018jco}
M.~Mezei, {{Membrane theory of entanglement dynamics from holography}},
  \href{http://dx.doi.org/10.1103/PhysRevD.98.106025}{Phys. Rev. D {\bf 98},
  106025, 2018},
  [\href{http://arxiv.org/abs/arXiv:1803.10244}{{arXiv:1803.10244 [hep-th]}}].

\bibitem{Mezei:2019zyt}
M.~Mezei and J.~Virrueta, {{Exploring the Membrane Theory of Entanglement
  Dynamics}}, \href{http://dx.doi.org/10.1007/JHEP02(2020)013}{JHEP {\bf 02},
  013, 2020}, [\href{http://arxiv.org/abs/arXiv:1912.11024}{{arXiv:1912.11024
  [hep-th]}}].

\bibitem{Liu:2020jsv}
H.~Liu and S.~Vardhan, {{Entanglement entropies of equilibrated pure states in
  quantum many-body systems and gravity}},
  \href{http://dx.doi.org/10.1103/PRXQuantum.2.010344}{P. R. X. Quantum. {\bf
  2}, 010344, 2021},
  [\href{http://arxiv.org/abs/arXiv:2008.01089}{{arXiv:2008.01089 [hep-th]}}].

\bibitem{Gao:2016bin}
P.~Gao, D.~L. Jafferis and A.~C. Wall, {{Traversable Wormholes via a Double
  Trace Deformation}}, \href{http://dx.doi.org/10.1007/JHEP12(2017)151}{JHEP
  {\bf 12}, 151, 2017},
  [\href{http://arxiv.org/abs/arXiv:1608.05687}{{arXiv:1608.05687 [hep-th]}}].

\bibitem{Gao:2018yzk}
P.~Gao and H.~Liu, {{Regenesis and quantum traversable wormholes}},
  \href{http://dx.doi.org/10.1007/JHEP10(2019)048}{JHEP {\bf 10}, 048, 2019},
  [\href{http://arxiv.org/abs/arXiv:1810.01444}{{arXiv:1810.01444 [hep-th]}}].

\bibitem{Murthy:2019fgs}
C.~Murthy and M.~Srednicki, {{Bounds on chaos from the eigenstate
  thermalization hypothesis}},
  \href{http://dx.doi.org/10.1103/PhysRevLett.123.230606}{Phys. Rev. Lett. {\bf
  123}, 230606, 2019},
  [\href{http://arxiv.org/abs/arXiv:1906.10808}{{arXiv:1906.10808
  [cond-mat.stat-mech]}}].

\bibitem{Tsuji:2017fxs}
N.~Tsuji, T.~Shitara and M.~Ueda, {{Bound on the exponential growth rate of
  out-of-time-ordered correlators}},
  \href{http://dx.doi.org/10.1103/PhysRevE.98.012216}{Phys. Rev. E {\bf 98},
  012216, 2018},
  [\href{http://arxiv.org/abs/arXiv:1706.09160}{{arXiv:1706.09160
  [cond-mat.stat-mech]}}].

\bibitem{Romero-Bermudez:2019vej}
A.~Romero-Berm\'udez, K.~Schalm and V.~Scopelliti, {{Regularization dependence
  of the OTOC. Which Lyapunov spectrum is the physical one?}},
  \href{http://dx.doi.org/10.1007/JHEP07(2019)107}{JHEP {\bf 07}, 107, 2019},
  [\href{http://arxiv.org/abs/arXiv:1903.09595}{{arXiv:1903.09595 [hep-th]}}].

\bibitem{Lucas:2018wsc}
A.~Lucas, {{Operator size at finite temperature and Planckian bounds on quantum
  dynamics}}, \href{http://dx.doi.org/10.1103/PhysRevLett.122.216601}{Phys.
  Rev. Lett. {\bf 122}, 216601, 2019},
  [\href{http://arxiv.org/abs/arXiv:1809.07769}{{arXiv:1809.07769
  [cond-mat.str-el]}}].

\bibitem{Qi:2018bje}
X.-L. Qi and A.~Streicher, {{Quantum Epidemiology: Operator Growth, Thermal
  Effects, and SYK}}, \href{http://dx.doi.org/10.1007/JHEP08(2019)012}{JHEP
  {\bf 08}, 012, 2019},
  [\href{http://arxiv.org/abs/arXiv:1810.11958}{{arXiv:1810.11958 [hep-th]}}].

\bibitem{Mousatov:2019xmc}
A.~Mousatov, {{Operator Size for Holographic Field Theories}},
  [\href{http://arxiv.org/abs/arXiv:1911.05089}{{arXiv:1911.05089 [hep-th]}}].

\bibitem{Lensky:2020ubw}
Y.~D. Lensky, X.-L. Qi and P.~Zhang, {{Size of bulk fermions in the SYK
  model}}, \href{http://dx.doi.org/10.1007/JHEP10(2020)053}{JHEP {\bf 10}, 053,
  2020}, [\href{http://arxiv.org/abs/arXiv:2002.01961}{{arXiv:2002.01961
  [hep-th]}}].

\bibitem{Brown:2018kvn}
A.~R. Brown, H.~Gharibyan, A.~Streicher, L.~Susskind, L.~Thorlacius and
  Y.~Zhao, {{Falling Toward Charged Black Holes}},
  \href{http://dx.doi.org/10.1103/PhysRevD.98.126016}{Phys. Rev. D {\bf 98},
  126016, 2018},
  [\href{http://arxiv.org/abs/arXiv:1804.04156}{{arXiv:1804.04156 [hep-th]}}].

\bibitem{Brown:2019hmk}
A.~R. Brown, H.~Gharibyan, S.~Leichenauer, H.~W. Lin, S.~Nezami, G.~Salton,
  L.~Susskind, B.~Swingle and M.~Walter, {{Quantum Gravity in the Lab:
  Teleportation by Size and Traversable Wormholes}},
  [\href{http://arxiv.org/abs/arXiv:1911.06314}{{arXiv:1911.06314
  [quant-ph]}}].

\bibitem{Nezami:2021yaq}
S.~Nezami, H.~W. Lin, A.~R. Brown, H.~Gharibyan, S.~Leichenauer, G.~Salton,
  L.~Susskind, B.~Swingle and M.~Walter, {{Quantum Gravity in the Lab:
  Teleportation by Size and Traversable Wormholes, Part II}},
  [\href{http://arxiv.org/abs/arXiv:2102.01064}{{arXiv:2102.01064
  [quant-ph]}}].

\bibitem{Schuster:2021uvg}
T.~Schuster, B.~Kobrin, P.~Gao, I.~Cong, E.~T. Khabiboulline, N.~M. Linke,
  M.~D. Lukin, C.~Monroe, B.~Yoshida and N.~Y. Yao, {{Many-body quantum
  teleportation via operator spreading in the traversable wormhole protocol}},
  [\href{http://arxiv.org/abs/arXiv:2102.00010}{{arXiv:2102.00010
  [quant-ph]}}].

\bibitem{Han:2018bqy}
X.~Han and S.~A. Hartnoll, {{Quantum Scrambling and State Dependence of the
  Butterfly Velocity}},
  \href{http://dx.doi.org/10.21468/SciPostPhys.7.4.045}{SciPost Phys. {\bf 7},
  045, 2019}, [\href{http://arxiv.org/abs/arXiv:1812.07598}{{arXiv:1812.07598
  [hep-th]}}].

\bibitem{Lucas:2019cxr}
A.~Lucas, {{Non-perturbative dynamics of the operator size distribution in the
  Sachdev\textendash{}Ye\textendash{}Kitaev model}},
  \href{http://dx.doi.org/10.1063/1.5133964}{J. Math. Phys. {\bf 61}, 081901,
  2020}, [\href{http://arxiv.org/abs/arXiv:1910.09539}{{arXiv:1910.09539
  [hep-th]}}].

\bibitem{Tran:2020xpc}
M.~C. Tran, C.-F. Chen, A.~Ehrenberg, A.~Y. Guo, A.~Deshpande, Y.~Hong, Z.-X.
  Gong, A.~V. Gorshkov and A.~Lucas, {{Hierarchy of Linear Light Cones with
  Long-Range Interactions}},
  \href{http://dx.doi.org/10.1103/PhysRevX.10.031009}{Phys. Rev. X {\bf 10},
  031009, 2020},
  [\href{http://arxiv.org/abs/arXiv:2001.11509}{{arXiv:2001.11509
  [quant-ph]}}].

\bibitem{Zaanen2004}
J.~Zaanen, {Why the temperature is high},
  \href{http://dx.doi.org/10.1038/430512a}{Nature {\bf 430}, 512--513, 2004}.

\bibitem{Zaanen:2018edk}
J.~Zaanen, {{Planckian dissipation, minimal viscosity and the transport in
  cuprate strange metals}},
  \href{http://dx.doi.org/10.21468/SciPostPhys.6.5.061}{SciPost Phys. {\bf 6},
  061, 2019}, [\href{http://arxiv.org/abs/arXiv:1807.10951}{{arXiv:1807.10951
  [cond-mat.str-el]}}].

\bibitem{Hartnoll:2021ydi}
S.~A. Hartnoll and A.~P. Mackenzie, {{Planckian Dissipation in Metals}},
  [\href{http://arxiv.org/abs/arXiv:2107.07802}{{arXiv:2107.07802
  [cond-mat.str-el]}}].

\bibitem{Hartnoll:2014lpa}
S.~A. Hartnoll, {{Theory of universal incoherent metallic transport}},
  \href{http://dx.doi.org/10.1038/nphys3174}{Nature Phys. {\bf 11}, 54, 2015},
  [\href{http://arxiv.org/abs/arXiv:1405.3651}{{arXiv:1405.3651
  [cond-mat.str-el]}}].

\bibitem{Bruin2013}
J.~A.~N. Bruin, H.~Sakai, R.~S. Perry and A.~P. Mackenzie, {Similarity of
  scattering rates in metals showing {$T$}-linear resistivity},
  \href{http://dx.doi.org/10.1126/science.1227612}{Science {\bf 339}, 804--807,
  2013},
  [\href{http://arxiv.org/abs/https://www.science.org/doi/pdf/10.1126/science.1227612}{{https://www.science.org/doi/pdf/10.1126/science.1227612}}].

\bibitem{Legros}
A.~Legros, S.~Benhabib, W.~Tabis, F.~Laliberté, M.~Dion, M.~Lizaire,
  B.~Vignolle, D.~Vignolles, H.~Raffy, Z.~Z. Li, P.~Auban-Senzier,
  N.~Doiron-Leyraud, P.~Fournier et~al., {Universal t-linear resistivity and
  planckian dissipation in overdoped cuprates},
  \href{http://dx.doi.org/10.1038/s41567-018-0334-2}{Nature Physics {\bf 15},
  142–147, 2018}.

\bibitem{Donos:2012js}
A.~Donos and S.~A. Hartnoll, {{Interaction-driven localization in holography}},
  \href{http://dx.doi.org/10.1038/nphys2701}{Nature Phys. {\bf 9}, 649--655,
  2013}, [\href{http://arxiv.org/abs/arXiv:1212.2998}{{arXiv:1212.2998
  [hep-th]}}].

\bibitem{Gouteraux:2014hca}
B.~Gout\'eraux, {{Charge transport in holography with momentum dissipation}},
  \href{http://dx.doi.org/10.1007/JHEP04(2014)181}{JHEP {\bf 04}, 181, 2014},
  [\href{http://arxiv.org/abs/arXiv:1401.5436}{{arXiv:1401.5436 [hep-th]}}].

\bibitem{Donos:2014uba}
A.~Donos and J.~P. Gauntlett, {{Novel metals and insulators from holography}},
  \href{http://dx.doi.org/10.1007/JHEP06(2014)007}{JHEP {\bf 06}, 007, 2014},
  [\href{http://arxiv.org/abs/arXiv:1401.5077}{{arXiv:1401.5077 [hep-th]}}].

\bibitem{Baggioli:2016oqk}
M.~Baggioli and O.~Pujolas, {{On holographic disorder-driven metal-insulator
  transitions}}, \href{http://dx.doi.org/10.1007/JHEP01(2017)040}{JHEP {\bf
  01}, 040, 2017},
  [\href{http://arxiv.org/abs/arXiv:1601.07897}{{arXiv:1601.07897 [hep-th]}}].

\bibitem{balents17}
X.-Y. Song, C.-M. Jian and L.~Balents, {Strongly correlated metal built from
  sachdev-ye-kitaev models},
  \href{http://dx.doi.org/10.1103/PhysRevLett.119.216601}{Phys. Rev. Lett. {\bf
  119}, 216601, 2017},
  [\href{http://arxiv.org/abs/arXiv:1705.00117}{{arXiv:1705.00117
  [cond-mat.str-el]}}].

\bibitem{Patel:2017mjv}
A.~A. Patel, J.~McGreevy, D.~P. Arovas and S.~Sachdev, {{Magnetotransport in a
  model of a disordered strange metal}},
  \href{http://dx.doi.org/10.1103/PhysRevX.8.021049}{Phys. Rev. X {\bf 8},
  021049, 2018},
  [\href{http://arxiv.org/abs/arXiv:1712.05026}{{arXiv:1712.05026
  [cond-mat.str-el]}}].

\bibitem{Chowdhury:2018sho}
D.~Chowdhury, Y.~Werman, E.~Berg and T.~Senthil, {{Translationally invariant
  non-Fermi liquid metals with critical Fermi-surfaces: Solvable models}},
  \href{http://dx.doi.org/10.1103/PhysRevX.8.031024}{Phys. Rev. X {\bf 8},
  031024, 2018},
  [\href{http://arxiv.org/abs/arXiv:1801.06178}{{arXiv:1801.06178
  [cond-mat.str-el]}}].

\bibitem{Patel:2019qce}
A.~A. Patel and S.~Sachdev, {{Theory of a Planckian metal}},
  \href{http://dx.doi.org/10.1103/PhysRevLett.123.066601}{Phys. Rev. Lett. {\bf
  123}, 066601, 2019},
  [\href{http://arxiv.org/abs/arXiv:1906.03265}{{arXiv:1906.03265
  [cond-mat.str-el]}}].

\bibitem{Baggioli:2016pia}
M.~Baggioli, B.~Gout\'eraux, E.~Kiritsis and W.-J. Li, {{Higher derivative
  corrections to incoherent metallic transport in holography}},
  \href{http://dx.doi.org/10.1007/JHEP03(2017)170}{JHEP {\bf 03}, 170, 2017},
  [\href{http://arxiv.org/abs/arXiv:1612.05500}{{arXiv:1612.05500 [hep-th]}}].

\bibitem{Guo:2019csw}
H.~Guo, Y.~Gu and S.~Sachdev, {{Transport and chaos in lattice
  Sachdev-Ye-Kitaev models}},
  \href{http://dx.doi.org/10.1103/PhysRevB.100.045140}{Phys. Rev. B {\bf 100},
  045140, 2019},
  [\href{http://arxiv.org/abs/arXiv:1904.02174}{{arXiv:1904.02174
  [cond-mat.str-el]}}].

\bibitem{Aleiner:2016eni}
I.~L. Aleiner, L.~Faoro and L.~B. Ioffe, {{Microscopic model of quantum
  butterfly effect: out-of-time-order correlators and traveling combustion
  waves}}, \href{http://dx.doi.org/10.1016/j.aop.2016.09.006}{Annals Phys. {\bf
  375}, 378--406, 2016},
  [\href{http://arxiv.org/abs/arXiv:1609.01251}{{arXiv:1609.01251
  [cond-mat.stat-mech]}}].

\bibitem{Grozdanov:2018atb}
S.~Grozdanov, K.~Schalm and V.~Scopelliti, {{Kinetic theory for classical and
  quantum many-body chaos}},
  \href{http://dx.doi.org/10.1103/PhysRevE.99.012206}{Phys. Rev. E {\bf 99},
  012206, 2019},
  [\href{http://arxiv.org/abs/arXiv:1804.09182}{{arXiv:1804.09182 [hep-th]}}].

\bibitem{Patel:2016wdy}
A.~A. Patel and S.~Sachdev, {{Quantum chaos on a critical Fermi surface}},
  \href{http://dx.doi.org/10.1073/pnas.1618185114}{Proc. Nat. Acad. Sci. {\bf
  114}, 1844--1849, 2017},
  [\href{http://arxiv.org/abs/arXiv:1611.00003}{{arXiv:1611.00003
  [cond-mat.str-el]}}].

\bibitem{Werman:2017abn}
Y.~Werman, S.~A. Kivelson and E.~Berg, {{Quantum chaos in an electron-phonon
  bad metal}},  [\href{http://arxiv.org/abs/arXiv:1705.07895}{{arXiv:1705.07895
  [cond-mat.str-el]}}].

\bibitem{Patel:2017vfp}
A.~A. Patel, D.~Chowdhury, S.~Sachdev and B.~Swingle, {{Quantum butterfly
  effect in weakly interacting diffusive metals}},
  \href{http://dx.doi.org/10.1103/PhysRevX.7.031047}{Phys. Rev. X {\bf 7},
  031047, 2017},
  [\href{http://arxiv.org/abs/arXiv:1703.07353}{{arXiv:1703.07353
  [cond-mat.str-el]}}].

\bibitem{Lucas:2016yfl}
A.~Lucas and J.~Steinberg, {{Charge diffusion and the butterfly effect in
  striped holographic matter}},
  \href{http://dx.doi.org/10.1007/JHEP10(2016)143}{JHEP {\bf 10}, 143, 2016},
  [\href{http://arxiv.org/abs/arXiv:1608.03286}{{arXiv:1608.03286 [hep-th]}}].

\bibitem{Hartman:2017hhp}
T.~Hartman, S.~A. Hartnoll and R.~Mahajan, {{Upper Bound on Diffusivity}},
  \href{http://dx.doi.org/10.1103/PhysRevLett.119.141601}{Phys. Rev. Lett. {\bf
  119}, 141601, 2017},
  [\href{http://arxiv.org/abs/arXiv:1706.00019}{{arXiv:1706.00019 [hep-th]}}].

\bibitem{Delacretaz:2021ufg}
L.~V. Delacretaz, A.~L. Fitzpatrick, E.~Katz and M.~T. Walters,
  {{Thermalization and Hydrodynamics of Two-Dimensional Quantum Field
  Theories}},  [\href{http://arxiv.org/abs/arXiv:2105.02229}{{arXiv:2105.02229
  [hep-th]}}].

\bibitem{Grozdanov:2020koi}
S.~Grozdanov, {{Bounds on transport from univalence and pole-skipping}},
  \href{http://dx.doi.org/10.1103/PhysRevLett.126.051601}{Phys. Rev. Lett. {\bf
  126}, 051601, 2021},
  [\href{http://arxiv.org/abs/arXiv:2008.00888}{{arXiv:2008.00888 [hep-th]}}].

\bibitem{Horowitz:1999jd}
G.~T. Horowitz and V.~E. Hubeny, {{Quasinormal modes of AdS black holes and the
  approach to thermal equilibrium}},
  \href{http://dx.doi.org/10.1103/PhysRevD.62.024027}{Phys. Rev. {\bf D62},
  024027, 2000},
  [\href{http://arxiv.org/abs/arXiv:hep-th/9909056}{{arXiv:hep-th/9909056
  [hep-th]}}].

\bibitem{Starinets:2002br}
A.~O. Starinets, {{Quasinormal modes of near extremal black branes}},
  \href{http://dx.doi.org/10.1103/PhysRevD.66.124013}{Phys. Rev. {\bf D66},
  124013, 2002},
  [\href{http://arxiv.org/abs/arXiv:hep-th/0207133}{{arXiv:hep-th/0207133
  [hep-th]}}].

\bibitem{Motl:2003cd}
L.~Motl and A.~Neitzke, {{Asymptotic black hole quasinormal frequencies}},
  \href{http://dx.doi.org/10.4310/ATMP.2003.v7.n2.a4}{Adv. Theor. Math. Phys.
  {\bf 7}, 307--330, 2003},
  [\href{http://arxiv.org/abs/arXiv:hep-th/0301173}{{arXiv:hep-th/0301173
  [hep-th]}}].

\bibitem{Kovtun:2005ev}
P.~K. Kovtun and A.~O. Starinets, {{Quasinormal modes and holography}},
  \href{http://dx.doi.org/10.1103/PhysRevD.72.086009}{Phys.Rev. {\bf D72},
  086009, 2005},
  [\href{http://arxiv.org/abs/arXiv:hep-th/0506184}{{arXiv:hep-th/0506184
  [hep-th]}}].

\bibitem{Festuccia:2008zx}
G.~Festuccia and H.~Liu, {{A Bohr-Sommerfeld quantization formula for
  quasinormal frequencies of AdS black holes}},
  \href{http://dx.doi.org/10.1166/asl.2009.1029}{Adv. Sci. Lett. {\bf 2},
  221--235, 2009},
  [\href{http://arxiv.org/abs/arXiv:0811.1033}{{arXiv:0811.1033 [gr-qc]}}].

\bibitem{Sybesma:2015oha}
W.~Sybesma and S.~Vandoren, {{Lifshitz quasinormal modes and relaxation from
  holography}}, \href{http://dx.doi.org/10.1007/JHEP05(2015)021}{JHEP {\bf 05},
  021, 2015}, [\href{http://arxiv.org/abs/arXiv:1503.07457}{{arXiv:1503.07457
  [hep-th]}}].

\bibitem{10.2307/1970079}
E.~P. Wigner, {Characteristic vectors of bordered matrices with infinite
  dimensions}, {Annals of Mathematics {\bf 62}, 548--564, 1955}.

\bibitem{PhysRevLett.52.1}
O.~Bohigas, M.~J. Giannoni and C.~Schmit, {Characterization of chaotic quantum
  spectra and universality of level fluctuation laws},
  \href{http://dx.doi.org/10.1103/PhysRevLett.52.1}{Phys. Rev. Lett. {\bf 52},
  1--4, 1984}.

\bibitem{GG_2016}
A.~M. Garc\'\i{}a-Garc\'\i{}a and J.~J.~M. Verbaarschot, {{Spectral and
  thermodynamic properties of the Sachdev-Ye-Kitaev model}},
  \href{http://dx.doi.org/10.1103/PhysRevD.94.126010}{Phys. Rev. D {\bf 94},
  126010, 2016},
  [\href{http://arxiv.org/abs/arXiv:1610.03816}{{arXiv:1610.03816 [hep-th]}}].

\bibitem{Bertini_2018}
B.~Bertini, P.~Kos and T.~Prosen, {{Exact Spectral Form Factor in a Minimal
  Model of Many-Body Quantum Chaos}},
  \href{http://dx.doi.org/10.1103/PhysRevLett.121.264101}{Phys. Rev. Lett. {\bf
  121}, 264101, 2018},
  [\href{http://arxiv.org/abs/arXiv:1805.00931}{{arXiv:1805.00931 [nlin.CD]}}].

\bibitem{Altland_2021}
A.~Altland and J.~Sonner, {{Late time physics of holographic quantum chaos}},
  \href{http://dx.doi.org/10.21468/SciPostPhys.11.2.034}{SciPost Phys. {\bf
  11}, 034, 2021},
  [\href{http://arxiv.org/abs/arXiv:2008.02271}{{arXiv:2008.02271 [hep-th]}}].

\bibitem{DAlessio_2016}
L.~D'Alessio, Y.~Kafri, A.~Polkovnikov and M.~Rigol, {{From quantum chaos and
  eigenstate thermalization to statistical mechanics and thermodynamics}},
  \href{http://dx.doi.org/10.1080/00018732.2016.1198134}{Adv. Phys. {\bf 65},
  239--362, 2016},
  [\href{http://arxiv.org/abs/arXiv:1509.06411}{{arXiv:1509.06411
  [cond-mat.stat-mech]}}].

\bibitem{HJ_2017}
J.~Cotler, N.~Hunter-Jones, J.~Liu and B.~Yoshida, {{Chaos, Complexity, and
  Random Matrices}}, \href{http://dx.doi.org/10.1007/JHEP11(2017)048}{JHEP {\bf
  11}, 048, 2017},
  [\href{http://arxiv.org/abs/arXiv:1706.05400}{{arXiv:1706.05400 [hep-th]}}].

\bibitem{susskind2018lectures}
L.~Susskind, {{Three Lectures on Complexity and Black Holes}},
  [\href{http://arxiv.org/abs/arXiv:1810.11563}{{arXiv:1810.11563 [hep-th]}}].

\bibitem{dorfman_1999}
J.~R. Dorfman, \emph{An Introduction to Chaos in Nonequilibrium Statistical
  Mechanics}.
\newblock Cambridge Lecture Notes in Physics. Cambridge University Press, 1999,
  \href{http://dx.doi.org/10.1017/CBO9780511628870}{10.1017/CBO9780511628870}.

\bibitem{Almheiri:2015aa}
A.~Almheiri, X.~Dong and D.~Harlow, {{Bulk Locality and Quantum Error
  Correction in AdS/CFT}},
  \href{http://dx.doi.org/10.1007/JHEP04(2015)163}{JHEP {\bf 04}, 163, 2015},
  [\href{http://arxiv.org/abs/arXiv:1411.7041}{{arXiv:1411.7041 [hep-th]}}].

\bibitem{Pastawski15b}
F.~Pastawski, B.~Yoshida, D.~Harlow and J.~Preskill, {{Holographic quantum
  error-correcting codes: Toy models for the bulk/boundary correspondence}},
  \href{http://dx.doi.org/10.1007/JHEP06(2015)149}{JHEP {\bf 06}, 149, 2015},
  [\href{http://arxiv.org/abs/arXiv:1503.06237}{{arXiv:1503.06237 [hep-th]}}].

\bibitem{Hayden:2016aa}
P.~Hayden, S.~Nezami, X.-L. Qi, N.~Thomas, M.~Walter and Z.~Yang, {{Holographic
  duality from random tensor networks}},
  \href{http://dx.doi.org/10.1007/JHEP11(2016)009}{JHEP {\bf 11}, 009, 2016},
  [\href{http://arxiv.org/abs/arXiv:1601.01694}{{arXiv:1601.01694 [hep-th]}}].

\bibitem{Dong2016}
X.~Dong, D.~Harlow and A.~C. Wall, {{Reconstruction of Bulk Operators within
  the Entanglement Wedge in Gauge-Gravity Duality}},
  \href{http://dx.doi.org/10.1103/PhysRevLett.117.021601}{Phys. Rev. Lett. {\bf
  117}, 021601, 2016},
  [\href{http://arxiv.org/abs/arXiv:1601.05416}{{arXiv:1601.05416 [hep-th]}}].

\bibitem{Almheiri19}
A.~Almheiri, N.~Engelhardt, D.~Marolf and H.~Maxfield, {{The entropy of bulk
  quantum fields and the entanglement wedge of an evaporating black hole}},
  \href{http://dx.doi.org/10.1007/JHEP12(2019)063}{JHEP {\bf 12}, 063, 2019},
  [\href{http://arxiv.org/abs/arXiv:1905.08762}{{arXiv:1905.08762 [hep-th]}}].

\bibitem{Penington19}
G.~Penington, {{Entanglement Wedge Reconstruction and the Information
  Paradox}}, \href{http://dx.doi.org/10.1007/JHEP09(2020)002}{JHEP {\bf 09},
  002, 2020}, [\href{http://arxiv.org/abs/arXiv:1905.08255}{{arXiv:1905.08255
  [hep-th]}}].

\bibitem{Hartman:2013qma}
T.~Hartman and J.~Maldacena, {{Time Evolution of Entanglement Entropy from
  Black Hole Interiors}}, \href{http://dx.doi.org/10.1007/JHEP05(2013)014}{JHEP
  {\bf 05}, 014, 2013},
  [\href{http://arxiv.org/abs/arXiv:1303.1080}{{arXiv:1303.1080 [hep-th]}}].

\bibitem{Stanford:2014aa}
D.~Stanford and L.~Susskind, {{Complexity and Shock Wave Geometries}},
  \href{http://dx.doi.org/10.1103/PhysRevD.90.126007}{Phys. Rev. D {\bf 90},
  126007, 2014}, [\href{http://arxiv.org/abs/arXiv:1406.2678}{{arXiv:1406.2678
  [hep-th]}}].

\bibitem{Brown19}
A.~R. Brown, H.~Gharibyan, G.~Penington and L.~Susskind, {{The
  Python\textquoteright{}s Lunch: geometric obstructions to decoding Hawking
  radiation}}, \href{http://dx.doi.org/10.1007/JHEP08(2020)121}{JHEP {\bf 08},
  121, 2020}, [\href{http://arxiv.org/abs/arXiv:1912.00228}{{arXiv:1912.00228
  [hep-th]}}].

\bibitem{Nahum:2017aa}
A.~Nahum, J.~Ruhman, S.~Vijay and J.~Haah, {{Quantum Entanglement Growth Under
  Random Unitary Dynamics}},
  \href{http://dx.doi.org/10.1103/PhysRevX.7.031016}{Phys. Rev. X {\bf 7},
  031016, 2017},
  [\href{http://arxiv.org/abs/arXiv:1608.06950}{{arXiv:1608.06950
  [cond-mat.stat-mech]}}].

\bibitem{Roberts:2017aa}
D.~A. Roberts and B.~Yoshida, {{Chaos and complexity by design}},
  \href{http://dx.doi.org/10.1007/JHEP04(2017)121}{JHEP {\bf 04}, 121, 2017},
  [\href{http://arxiv.org/abs/arXiv:1610.04903}{{arXiv:1610.04903
  [quant-ph]}}].

\bibitem{Brandao19}
F.~G. S.~L. Brand\~ao, W.~Chemissany, N.~Hunter-Jones, R.~Kueng and
  J.~Preskill, {{Models of Quantum Complexity Growth}},
  \href{http://dx.doi.org/10.1103/PRXQuantum.2.030316}{PRX Quantum {\bf 2},
  030316, 2021},
  [\href{http://arxiv.org/abs/arXiv:1912.04297}{{arXiv:1912.04297 [hep-th]}}].

\bibitem{Haferkamp21}
J.~Haferkamp, P.~Faist, N.~B.~T. Kothakonda, J.~Eisert and N.~Y. Halpern,
  {{Linear growth of quantum circuit complexity}},
  [\href{http://arxiv.org/abs/arXiv:2106.05305}{{arXiv:2106.05305
  [quant-ph]}}].

\bibitem{HP2007}
P.~Hayden and J.~Preskill, {{Black holes as mirrors: Quantum information in
  random subsystems}},
  \href{http://dx.doi.org/10.1088/1126-6708/2007/09/120}{JHEP {\bf 09}, 120,
  2007}, [\href{http://arxiv.org/abs/arXiv:0708.4025}{{arXiv:0708.4025
  [hep-th]}}].

\bibitem{Hosur:2015ylk}
P.~Hosur, X.-L. Qi, D.~A. Roberts and B.~Yoshida, {{Chaos in quantum
  channels}}, \href{http://dx.doi.org/10.1007/JHEP02(2016)004}{JHEP {\bf 02},
  004, 2016}, [\href{http://arxiv.org/abs/arXiv:1511.04021}{{arXiv:1511.04021
  [hep-th]}}].

\bibitem{Yoshida:2017aa}
B.~Yoshida and A.~Kitaev, {{Efficient decoding for the Hayden-Preskill
  protocol}},  [\href{http://arxiv.org/abs/arXiv:1710.03363}{{arXiv:1710.03363
  [hep-th]}}].

\bibitem{Landsman:2019aa}
K.~A. Landsman, C.~Figgatt, T.~Schuster, N.~M. Linke, B.~Yoshida, N.~Y. Yao and
  C.~Monroe, {{Verified Quantum Information Scrambling}},
  \href{http://dx.doi.org/10.1038/s41586-019-0952-6}{Nature {\bf 567}, 61--65,
  2019}, [\href{http://arxiv.org/abs/arXiv:1806.02807}{{arXiv:1806.02807
  [quant-ph]}}].

\bibitem{May:2019aa}
A.~May, {{Quantum tasks in holography}},
  \href{http://dx.doi.org/10.1007/JHEP10(2019)233}{JHEP {\bf 10}, 233, 2019},
  [\href{http://arxiv.org/abs/arXiv:1902.06845}{{arXiv:1902.06845 [hep-th]}}].

\bibitem{Vaidman:2003aa}
L.~Vaidman, {Instantaneous measurement of nonlocal variables},
  \href{http://dx.doi.org/10.1103/PhysRevLett.90.010402}{Phys. Rev. Lett. {\bf
  90}, 010402, 2003},
  [\href{http://arxiv.org/abs/arXiv:quant-ph/0111124}{{arXiv:quant-ph/0111124
  [quant-ph]}}].

\bibitem{Beigi:2011aa}
S.~Beigi and R.~K{\"o}nig, {Simplified instantaneous non-local quantum
  computation with applications to position-based cryptography},
  \href{http://dx.doi.org/10.1088/1367-2630/13/9/093036}{New. J. Phys. {\bf
  13}, 093036, 2011},
  [\href{http://arxiv.org/abs/arXiv:1101.1065}{{arXiv:1101.1065 [quant-ph]}}].

\bibitem{May:2020aa}
A.~May, G.~Penington and J.~Sorce, {{Holographic scattering requires a
  connected entanglement wedge}},
  \href{http://dx.doi.org/10.1007/JHEP08(2020)132}{JHEP {\bf 08}, 132, 2020},
  [\href{http://arxiv.org/abs/arXiv:1912.05649}{{arXiv:1912.05649 [hep-th]}}].

\bibitem{chowdhury2021sachdev}
D.~Chowdhury, A.~Georges, O.~Parcollet and S.~Sachdev, {{Sachdev-Ye-Kitaev
  Models and Beyond: A Window into Non-Fermi Liquids}},
  [\href{http://arxiv.org/abs/arXiv:2109.05037}{{arXiv:2109.05037
  [cond-mat.str-el]}}].

\bibitem{si2001locally}
Q.~{Si}, S.~{Rabello}, K.~{Ingersent} and J.~L. {Smith}, {{Locally critical
  quantum phase transitions in strongly correlated metals}},
  \href{http://dx.doi.org/10.1038/35101507}{Nature {\bf 413}, 804--808, 2001},
  [\href{http://arxiv.org/abs/arXiv:cond-mat/0011477}{{arXiv:cond-mat/0011477
  [cond-mat.str-el]}}].

\bibitem{Faulkner:2011tm}
T.~Faulkner, N.~Iqbal, H.~Liu, J.~McGreevy and D.~Vegh, {{Holographic non-Fermi
  liquid fixed points}}, \href{http://dx.doi.org/10.1098/rsta.2010.0354}{Phil.
  Trans. Roy. Soc. {\bf A 369}, 1640, 2011},
  [\href{http://arxiv.org/abs/arXiv:1101.0597}{{arXiv:1101.0597 [hep-th]}}].

\bibitem{J1984}
R.~Jackiw, {{Lower Dimensional Gravity}},
  \href{http://dx.doi.org/10.1016/0550-3213(85)90448-1}{Nucl. Phys. B {\bf
  252}, 343--356, 1985}.

\bibitem{T1983}
C.~Teitelboim, {{Gravitation and Hamiltonian Structure in Two Space-Time
  Dimensions}}, \href{http://dx.doi.org/10.1016/0370-2693(83)90012-6}{Phys.
  Lett. B {\bf 126}, 41--45, 1983}.

\bibitem{Saad:2019lba}
P.~Saad, S.~H. Shenker and D.~Stanford, {{JT gravity as a matrix integral}},
  [\href{http://arxiv.org/abs/arXiv:1903.11115}{{arXiv:1903.11115 [hep-th]}}].

\bibitem{Penington:2019kki}
G.~Penington, S.~H. Shenker, D.~Stanford and Z.~Yang, {{Replica wormholes and
  the black hole interior}},
  [\href{http://arxiv.org/abs/arXiv:1911.11977}{{arXiv:1911.11977 [hep-th]}}].

\bibitem{Almheiri:2019qdq}
A.~Almheiri, T.~Hartman, J.~Maldacena, E.~Shaghoulian and A.~Tajdini, {{Replica
  Wormholes and the Entropy of Hawking Radiation}},
  \href{http://dx.doi.org/10.1007/JHEP05(2020)013}{JHEP {\bf 05}, 013, 2020},
  [\href{http://arxiv.org/abs/arXiv:1911.12333}{{arXiv:1911.12333 [hep-th]}}].

\bibitem{Gu:2017ohj}
Y.~Gu, A.~Lucas and X.-L. Qi, {{Energy diffusion and the butterfly effect in
  inhomogeneous Sachdev-Ye-Kitaev chains}},
  \href{http://dx.doi.org/10.21468/SciPostPhys.2.3.018}{SciPost Phys. {\bf 2},
  018, 2017}, [\href{http://arxiv.org/abs/arXiv:1702.08462}{{arXiv:1702.08462
  [hep-th]}}].

\bibitem{GKZ21}
Y.~Gu, A.~Kitaev and P.~Zhang, {{A two-way approach to out-of-time-order
  correlators}},
  [\href{http://arxiv.org/abs/arXiv:2111.12007}{{arXiv:2111.12007 [hep-th]}}].

\bibitem{lin2018out}
C.-J. {Lin} and O.~I. {Motrunich}, {{Out-of-time-ordered correlators in a
  quantum Ising chain}},
  \href{http://dx.doi.org/10.1103/PhysRevB.97.144304}{PRB {\bf 97}, 144304,
  2018}, [\href{http://arxiv.org/abs/arXiv:1801.01636}{{arXiv:1801.01636
  [cond-mat.stat-mech]}}].

\bibitem{xu2020accessing}
S.~Xu and B.~Swingle, {{Accessing scrambling using matrix product operators}},
  \href{http://dx.doi.org/10.1038/s41567-019-0712-4}{Nature Phys. {\bf 16},
  199--204, 2019},
  [\href{http://arxiv.org/abs/arXiv:1802.00801}{{arXiv:1802.00801
  [quant-ph]}}].

\bibitem{Heemskerk:2009pn}
I.~Heemskerk, J.~Penedones, J.~Polchinski and J.~Sully, {{Holography from
  Conformal Field Theory}},
  \href{http://dx.doi.org/10.1088/1126-6708/2009/10/079}{JHEP {\bf 10}, 079,
  2009}, [\href{http://arxiv.org/abs/arXiv:0907.0151}{{arXiv:0907.0151
  [hep-th]}}].

\bibitem{ZGK20}
P.~Zhang, Y.~Gu and A.~Kitaev, {{An obstacle to sub-AdS holography for SYK-like
  models}}, \href{http://dx.doi.org/10.1007/JHEP03(2021)094}{JHEP {\bf 21},
  094, 2020}, [\href{http://arxiv.org/abs/arXiv:2012.01620}{{arXiv:2012.01620
  [hep-th]}}].

\bibitem{franz2018mimicking}
M.~Franz and M.~Rozali, {{Mimicking black hole event horizons in atomic and
  solid-state systems}},
  \href{http://dx.doi.org/10.1038/s41578-018-0058-z}{Nature Rev. Mater. {\bf
  3}, 491--501, 2018},
  [\href{http://arxiv.org/abs/arXiv:1808.00541}{{arXiv:1808.00541
  [cond-mat.str-el]}}].

\bibitem{Blok:2020may}
M.~S. Blok, V.~V. Ramasesh, T.~Schuster, K.~O'Brien, J.~M. Kreikebaum,
  D.~Dahlen, A.~Morvan, B.~Yoshida, N.~Y. Yao and I.~Siddiqi, {{Quantum
  Information Scrambling on a Superconducting Qutrit Processor}},
  \href{http://dx.doi.org/10.1103/PhysRevX.11.021010}{Phys. Rev. X {\bf 11},
  021010, 2021},
  [\href{http://arxiv.org/abs/arXiv:2003.03307}{{arXiv:2003.03307
  [quant-ph]}}].

\bibitem{Schafer:2009dj}
T.~Sch\"afer and D.~Teaney, {{Nearly Perfect Fluidity: From Cold Atomic Gases
  to Hot Quark Gluon Plasmas}},
  \href{http://dx.doi.org/10.1088/0034-4885/72/12/126001}{Rept. Prog. Phys.
  {\bf 72}, 126001, 2009},
  [\href{http://arxiv.org/abs/arXiv:0904.3107}{{arXiv:0904.3107 [hep-ph]}}].

\bibitem{Adams:2012th}
A.~Adams, L.~D. Carr, T.~Sch\"afer, P.~Steinberg and J.~E. Thomas, {{Strongly
  Correlated Quantum Fluids: Ultracold Quantum Gases, Quantum Chromodynamic
  Plasmas, and Holographic Duality}},
  \href{http://dx.doi.org/10.1088/1367-2630/14/11/115009}{New J. Phys. {\bf
  14}, 115009, 2012},
  [\href{http://arxiv.org/abs/arXiv:1205.5180}{{arXiv:1205.5180 [hep-th]}}].

\bibitem{Jacak:2012dx}
B.~V. Jacak and B.~Muller, {{The exploration of hot nuclear matter}},
  \href{http://dx.doi.org/10.1126/science.1215901}{Science {\bf 337}, 310--314,
  2012}.

\bibitem{dwdw}
O.~DeWolfe, S.~S. Gubser, C.~Rosen and D.~Teaney, {{Heavy ions and string
  theory}}, \href{http://dx.doi.org/10.1016/j.ppnp.2013.11.001}{Prog. Part.
  Nucl. Phys. {\bf 75}, 86--132, 2014},
  [\href{http://arxiv.org/abs/arXiv:1304.7794}{{arXiv:1304.7794 [hep-th]}}].

\bibitem{Romatschke:2017ejr}
P.~Romatschke and U.~Romatschke, \emph{{Relativistic Fluid Dynamics In and Out
  of Equilibrium}}.
\newblock Cambridge Monographs on Mathematical Physics. Cambridge University
  Press, 2019,
  \href{http://dx.doi.org/10.1017/9781108651998}{10.1017/9781108651998}.

\bibitem{Busza:2018rrf}
W.~Busza, K.~Rajagopal and W.~van~der Schee, {{Heavy Ion Collisions: The Big
  Picture, and the Big Questions}},
  \href{http://dx.doi.org/10.1146/annurev-nucl-101917-020852}{Ann. Rev. Nucl.
  Part. Sci. {\bf 68}, 339--376, 2018},
  [\href{http://arxiv.org/abs/arXiv:1802.04801}{{arXiv:1802.04801 [hep-ph]}}].

\bibitem{Gubser:1996de}
S.~S. Gubser, I.~R. Klebanov and A.~W. Peet, {{Entropy and temperature of black
  3-branes}}, \href{http://dx.doi.org/10.1103/PhysRevD.54.3915}{Phys. Rev. D
  {\bf 54}, 3915--3919, 1996},
  [\href{http://arxiv.org/abs/arXiv:hep-th/9602135}{{arXiv:hep-th/9602135}}].

\bibitem{Witten:1998zw}
E.~Witten, {{Anti-de Sitter space, thermal phase transition, and confinement in
  gauge theories}}, \href{http://dx.doi.org/10.4310/ATMP.1998.v2.n3.a3}{Adv.
  Theor. Math. Phys. {\bf 2}, 505--532, 1998},
  [\href{http://arxiv.org/abs/arXiv:hep-th/9803131}{{arXiv:hep-th/9803131}}].

\bibitem{Borsanyi:2013bia}
S.~Borsanyi, Z.~Fodor, C.~Hoelbling, S.~D. Katz, S.~Krieg and K.~K. Szabo,
  {{Full result for the QCD equation of state with 2+1 flavors}},
  \href{http://dx.doi.org/10.1016/j.physletb.2014.01.007}{Phys. Lett. B {\bf
  730}, 99--104, 2014},
  [\href{http://arxiv.org/abs/arXiv:1309.5258}{{arXiv:1309.5258 [hep-lat]}}].

\bibitem{HotQCD:2014kol}
{\scshape HotQCD} collaboration, A.~Bazavov et~al., {{Equation of state in (
  2+1 )-flavor QCD}}, \href{http://dx.doi.org/10.1103/PhysRevD.90.094503}{Phys.
  Rev. D {\bf 90}, 094503, 2014},
  [\href{http://arxiv.org/abs/arXiv:1407.6387}{{arXiv:1407.6387 [hep-lat]}}].

\bibitem{Policastro:2002se}
G.~Policastro, D.~T. Son and A.~O. Starinets, {{From AdS / CFT correspondence
  to hydrodynamics}},
  \href{http://dx.doi.org/10.1088/1126-6708/2002/09/043}{JHEP {\bf 09}, 043,
  2002},
  [\href{http://arxiv.org/abs/arXiv:hep-th/0205052}{{arXiv:hep-th/0205052}}].

\bibitem{Son:2002sd}
D.~T. Son and A.~O. Starinets, {{Minkowski space correlators in AdS / CFT
  correspondence: Recipe and applications}},
  \href{http://dx.doi.org/10.1088/1126-6708/2002/09/042}{JHEP {\bf 09}, 042,
  2002},
  [\href{http://arxiv.org/abs/arXiv:hep-th/0205051}{{arXiv:hep-th/0205051}}].

\bibitem{Teaney:2003kp}
D.~Teaney, {{The Effects of viscosity on spectra, elliptic flow, and HBT
  radii}}, \href{http://dx.doi.org/10.1103/PhysRevC.68.034913}{Phys. Rev. C
  {\bf 68}, 034913, 2003},
  [\href{http://arxiv.org/abs/arXiv:nucl-th/0301099}{{arXiv:nucl-th/0301099}}].

\bibitem{Romatschke:2007mq}
P.~Romatschke and U.~Romatschke, {{Viscosity Information from Relativistic
  Nuclear Collisions: How Perfect is the Fluid Observed at RHIC?}},
  \href{http://dx.doi.org/10.1103/PhysRevLett.99.172301}{Phys. Rev. Lett. {\bf
  99}, 172301, 2007},
  [\href{http://arxiv.org/abs/arXiv:0706.1522}{{arXiv:0706.1522 [nucl-th]}}].

\bibitem{Luzum:2008cw}
M.~Luzum and P.~Romatschke, {{Conformal Relativistic Viscous Hydrodynamics:
  Applications to RHIC results at s(NN)**(1/2) = 200-GeV}},
  \href{http://dx.doi.org/10.1103/PhysRevC.78.034915}{Phys. Rev. C {\bf 78},
  034915, 2008}, [\href{http://arxiv.org/abs/arXiv:0804.4015}{{arXiv:0804.4015
  [nucl-th]}}].

\bibitem{Song:2010mg}
H.~Song, S.~A. Bass, U.~Heinz, T.~Hirano and C.~Shen, {{200 A GeV Au+Au
  collisions serve a nearly perfect quark-gluon liquid}},
  \href{http://dx.doi.org/10.1103/PhysRevLett.106.192301}{Phys. Rev. Lett. {\bf
  106}, 192301, 2011},
  [\href{http://arxiv.org/abs/arXiv:1011.2783}{{arXiv:1011.2783 [nucl-th]}}].

\bibitem{Heinz:2013th}
U.~Heinz and R.~Snellings, {{Collective flow and viscosity in relativistic
  heavy-ion collisions}},
  \href{http://dx.doi.org/10.1146/annurev-nucl-102212-170540}{Ann. Rev. Nucl.
  Part. Sci. {\bf 63}, 123--151, 2013},
  [\href{http://arxiv.org/abs/arXiv:1301.2826}{{arXiv:1301.2826 [nucl-th]}}].

\bibitem{Herzog:2006gh}
C.~P. Herzog, A.~Karch, P.~Kovtun, C.~Kozcaz and L.~G. Yaffe, {{Energy loss of
  a heavy quark moving through N=4 supersymmetric Yang-Mills plasma}},
  \href{http://dx.doi.org/10.1088/1126-6708/2006/07/013}{JHEP {\bf 07}, 013,
  2006},
  [\href{http://arxiv.org/abs/arXiv:hep-th/0605158}{{arXiv:hep-th/0605158}}].

\bibitem{Gubser:2006bz}
S.~S. Gubser, {{Drag force in AdS/CFT}},
  \href{http://dx.doi.org/10.1103/PhysRevD.74.126005}{Phys. Rev. D {\bf 74},
  126005, 2006},
  [\href{http://arxiv.org/abs/arXiv:hep-th/0605182}{{arXiv:hep-th/0605182}}].

\bibitem{Casalderrey-Solana:2006fio}
J.~Casalderrey-Solana and D.~Teaney, {{Heavy quark diffusion in strongly
  coupled N=4 Yang-Mills}},
  \href{http://dx.doi.org/10.1103/PhysRevD.74.085012}{Phys. Rev. D {\bf 74},
  085012, 2006},
  [\href{http://arxiv.org/abs/arXiv:hep-ph/0605199}{{arXiv:hep-ph/0605199}}].

\bibitem{Gubser:2006nz}
S.~S. Gubser, {{Momentum fluctuations of heavy quarks in the gauge-string
  duality}}, \href{http://dx.doi.org/10.1016/j.nuclphysb.2007.09.017}{Nucl.
  Phys. B {\bf 790}, 175--199, 2008},
  [\href{http://arxiv.org/abs/arXiv:hep-th/0612143}{{arXiv:hep-th/0612143}}].

\bibitem{Casalderrey-Solana:2007ahi}
J.~Casalderrey-Solana and D.~Teaney, {{Transverse Momentum Broadening of a Fast
  Quark in a N=4 Yang Mills Plasma}},
  \href{http://dx.doi.org/10.1088/1126-6708/2007/04/039}{JHEP {\bf 04}, 039,
  2007},
  [\href{http://arxiv.org/abs/arXiv:hep-th/0701123}{{arXiv:hep-th/0701123}}].

\bibitem{Reiten:2019fta}
J.~Reiten and A.~V. Sadofyev, {{Drag force to all orders in gradients}},
  \href{http://dx.doi.org/10.1007/JHEP07(2020)146}{JHEP {\bf 07}, 146, 2020},
  [\href{http://arxiv.org/abs/arXiv:1912.08816}{{arXiv:1912.08816 [hep-th]}}].

\bibitem{Liu:2006ug}
H.~Liu, K.~Rajagopal and U.~A. Wiedemann, {{Calculating the jet quenching
  parameter from AdS/CFT}},
  \href{http://dx.doi.org/10.1103/PhysRevLett.97.182301}{Phys. Rev. Lett. {\bf
  97}, 182301, 2006},
  [\href{http://arxiv.org/abs/arXiv:hep-ph/0605178}{{arXiv:hep-ph/0605178}}].

\bibitem{DEramo:2010wup}
F.~D'Eramo, H.~Liu and K.~Rajagopal, {{Transverse Momentum Broadening and the
  Jet Quenching Parameter, Redux}},
  \href{http://dx.doi.org/10.1103/PhysRevD.84.065015}{Phys. Rev. D {\bf 84},
  065015, 2011}, [\href{http://arxiv.org/abs/arXiv:1006.1367}{{arXiv:1006.1367
  [hep-ph]}}].

\bibitem{Chesler:2007an}
P.~M. Chesler and L.~G. Yaffe, {{The Wake of a quark moving through a
  strongly-coupled plasma}},
  \href{http://dx.doi.org/10.1103/PhysRevLett.99.152001}{Phys. Rev. Lett. {\bf
  99}, 152001, 2007},
  [\href{http://arxiv.org/abs/arXiv:0706.0368}{{arXiv:0706.0368 [hep-th]}}].

\bibitem{Chesler:2014jva}
P.~M. Chesler and K.~Rajagopal, {{Jet quenching in strongly coupled plasma}},
  \href{http://dx.doi.org/10.1103/PhysRevD.90.025033}{Phys. Rev. D {\bf 90},
  025033, 2014}, [\href{http://arxiv.org/abs/arXiv:1402.6756}{{arXiv:1402.6756
  [hep-th]}}].

\bibitem{Chesler:2015nqz}
P.~M. Chesler and K.~Rajagopal, {{On the Evolution of Jet Energy and Opening
  Angle in Strongly Coupled Plasma}},
  \href{http://dx.doi.org/10.1007/JHEP05(2016)098}{JHEP {\bf 05}, 098, 2016},
  [\href{http://arxiv.org/abs/arXiv:1511.07567}{{arXiv:1511.07567 [hep-th]}}].

\bibitem{Casalderrey-Solana:2014bpa}
J.~Casalderrey-Solana, D.~C. Gulhan, J.~G. Milhano, D.~Pablos and K.~Rajagopal,
  {{A Hybrid Strong/Weak Coupling Approach to Jet Quenching}},
  \href{http://dx.doi.org/10.1007/JHEP09(2015)175}{JHEP {\bf 10}, 019, 2014},
  [\href{http://arxiv.org/abs/arXiv:1405.3864}{{arXiv:1405.3864 [hep-ph]}}].

\bibitem{Casalderrey-Solana:2015vaa}
J.~Casalderrey-Solana, D.~C. Gulhan, J.~G. Milhano, D.~Pablos and K.~Rajagopal,
  {{Predictions for Boson-Jet Observables and Fragmentation Function Ratios
  from a Hybrid Strong/Weak Coupling Model for Jet Quenching}},
  \href{http://dx.doi.org/10.1007/JHEP03(2016)053}{JHEP {\bf 03}, 053, 2016},
  [\href{http://arxiv.org/abs/arXiv:1508.00815}{{arXiv:1508.00815 [hep-ph]}}].

\bibitem{Casalderrey-Solana:2016jvj}
J.~Casalderrey-Solana, D.~Gulhan, G.~Milhano, D.~Pablos and K.~Rajagopal,
  {{Angular Structure of Jet Quenching Within a Hybrid Strong/Weak Coupling
  Model}}, \href{http://dx.doi.org/10.1007/JHEP03(2017)135}{JHEP {\bf 03}, 135,
  2017}, [\href{http://arxiv.org/abs/arXiv:1609.05842}{{arXiv:1609.05842
  [hep-ph]}}].

\bibitem{Hulcher:2017cpt}
Z.~Hulcher, D.~Pablos and K.~Rajagopal, {{Resolution Effects in the Hybrid
  Strong/Weak Coupling Model}},
  \href{http://dx.doi.org/10.1007/JHEP03(2018)010}{JHEP {\bf 03}, 010, 2018},
  [\href{http://arxiv.org/abs/arXiv:1707.05245}{{arXiv:1707.05245 [hep-ph]}}].

\bibitem{Casalderrey-Solana:2018wrw}
J.~Casalderrey-Solana, Z.~Hulcher, G.~Milhano, D.~Pablos and K.~Rajagopal,
  {{Simultaneous description of hadron and jet suppression in heavy-ion
  collisions}}, \href{http://dx.doi.org/10.1103/PhysRevC.99.051901}{Phys. Rev.
  C {\bf 99}, 051901, 2019},
  [\href{http://arxiv.org/abs/arXiv:1808.07386}{{arXiv:1808.07386 [hep-ph]}}].

\bibitem{Casalderrey-Solana:2019ubu}
J.~Casalderrey-Solana, G.~Milhano, D.~Pablos and K.~Rajagopal, {{Modification
  of Jet Substructure in Heavy Ion Collisions as a Probe of the Resolution
  Length of Quark-Gluon Plasma}},
  \href{http://dx.doi.org/10.1007/JHEP01(2020)044}{JHEP {\bf 01}, 044, 2020},
  [\href{http://arxiv.org/abs/arXiv:1907.11248}{{arXiv:1907.11248 [hep-ph]}}].

\bibitem{Casalderrey-Solana:2020rsj}
J.~Casalderrey-Solana, J.~G. Milhano, D.~Pablos, K.~Rajagopal and X.~Yao, {{Jet
  Wake from Linearized Hydrodynamics}},
  \href{http://dx.doi.org/10.1007/JHEP05(2021)230}{JHEP {\bf 05}, 230, 2021},
  [\href{http://arxiv.org/abs/arXiv:2010.01140}{{arXiv:2010.01140 [hep-ph]}}].

\bibitem{Chesler:2010bi}
P.~M. Chesler and L.~G. Yaffe, {{Holography and colliding gravitational shock
  waves in asymptotically AdS$_{5}$ spacetime}},
  \href{http://dx.doi.org/10.1103/PhysRevLett.106.021601}{Phys. Rev. Lett. {\bf
  106}, 021601, 2011},
  [\href{http://arxiv.org/abs/arXiv:1011.3562}{{arXiv:1011.3562 [hep-th]}}].

\bibitem{Chesler:2013lia}
P.~M. Chesler and L.~G. Yaffe, {{Numerical solution of gravitational dynamics
  in asymptotically anti-de Sitter spacetimes}},
  \href{http://dx.doi.org/10.1007/JHEP07(2014)086}{JHEP {\bf 07}, 086, 2014},
  [\href{http://arxiv.org/abs/arXiv:1309.1439}{{arXiv:1309.1439 [hep-th]}}].

\bibitem{Casalderrey-Solana:2013aba}
J.~Casalderrey-Solana, M.~P. Heller, D.~Mateos and W.~van~der Schee, {{From
  full stopping to transparency in a holographic model of heavy ion
  collisions}}, \href{http://dx.doi.org/10.1103/PhysRevLett.111.181601}{Phys.
  Rev. Lett. {\bf 111}, 181601, 2013},
  [\href{http://arxiv.org/abs/arXiv:1305.4919}{{arXiv:1305.4919 [hep-th]}}].

\bibitem{vanderSchee:2013pia}
W.~van~der Schee, P.~Romatschke and S.~Pratt, {{Fully Dynamical Simulation of
  Central Nuclear Collisions}},
  \href{http://dx.doi.org/10.1103/PhysRevLett.111.222302}{Phys. Rev. Lett. {\bf
  111}, 222302, 2013},
  [\href{http://arxiv.org/abs/arXiv:1307.2539}{{arXiv:1307.2539 [nucl-th]}}].

\bibitem{Chesler:2016ceu}
P.~M. Chesler, {{How big are the smallest drops of quark-gluon plasma?}},
  \href{http://dx.doi.org/10.1007/JHEP03(2016)146}{JHEP {\bf 03}, 146, 2016},
  [\href{http://arxiv.org/abs/arXiv:1601.01583}{{arXiv:1601.01583 [hep-th]}}].

\bibitem{Grozdanov:2016zjj}
S.~Grozdanov and W.~van~der Schee, {{Coupling Constant Corrections in a
  Holographic Model of Heavy Ion Collisions}},
  \href{http://dx.doi.org/10.1103/PhysRevLett.119.011601}{Phys. Rev. Lett. {\bf
  119}, 011601, 2017},
  [\href{http://arxiv.org/abs/arXiv:1610.08976}{{arXiv:1610.08976 [hep-th]}}].

\bibitem{Bantilan:2018vjv}
H.~Bantilan, T.~Ishii and P.~Romatschke, {{Holographic Heavy-Ion Collisions:
  Analytic Solutions with Longitudinal Flow, Elliptic Flow and Vorticity}},
  \href{http://dx.doi.org/10.1016/j.physletb.2018.08.038}{Phys. Lett. B {\bf
  785}, 201--206, 2018},
  [\href{http://arxiv.org/abs/arXiv:1803.10774}{{arXiv:1803.10774 [nucl-th]}}].

\bibitem{Folkestad:2019lam}
A.~Folkestad, S.~Grozdanov, K.~Rajagopal and W.~van~der Schee, {{Coupling
  Constant Corrections in a Holographic Model of Heavy Ion Collisions with
  Nonzero Baryon Number Density}},
  \href{http://dx.doi.org/10.1007/JHEP12(2019)093}{JHEP {\bf 12}, 093, 2019},
  [\href{http://arxiv.org/abs/arXiv:1907.13134}{{arXiv:1907.13134 [hep-th]}}].

\bibitem{Chesler:2008hg}
P.~M. Chesler and L.~G. Yaffe, {{Horizon formation and far-from-equilibrium
  isotropization in supersymmetric Yang-Mills plasma}},
  \href{http://dx.doi.org/10.1103/PhysRevLett.102.211601}{Phys. Rev. Lett. {\bf
  102}, 211601, 2009},
  [\href{http://arxiv.org/abs/arXiv:0812.2053}{{arXiv:0812.2053 [hep-th]}}].

\bibitem{Chesler:2009cy}
P.~M. Chesler and L.~G. Yaffe, {{Boost invariant flow, black hole formation,
  and far-from-equilibrium dynamics in N = 4 supersymmetric Yang-Mills
  theory}}, \href{http://dx.doi.org/10.1103/PhysRevD.82.026006}{Phys. Rev. D
  {\bf 82}, 026006, 2010},
  [\href{http://arxiv.org/abs/arXiv:0906.4426}{{arXiv:0906.4426 [hep-th]}}].

\bibitem{Heller:2011ju}
M.~P. Heller, R.~A. Janik and P.~Witaszczyk, {{The characteristics of
  thermalization of boost-invariant plasma from holography}},
  \href{http://dx.doi.org/10.1103/PhysRevLett.108.201602}{Phys. Rev. Lett. {\bf
  108}, 201602, 2012},
  [\href{http://arxiv.org/abs/arXiv:1103.3452}{{arXiv:1103.3452 [hep-th]}}].

\bibitem{Keegan:2015avk}
L.~Keegan, A.~Kurkela, P.~Romatschke, W.~van~der Schee and Y.~Zhu, {{Weak and
  strong coupling equilibration in nonabelian gauge theories}},
  \href{http://dx.doi.org/10.1007/JHEP04(2016)031}{JHEP {\bf 04}, 031, 2016},
  [\href{http://arxiv.org/abs/arXiv:1512.05347}{{arXiv:1512.05347 [hep-th]}}].

\bibitem{Heller:2016gbp}
M.~P. Heller, {{Holography, Hydrodynamization and Heavy-Ion Collisions}},
  \href{http://dx.doi.org/10.5506/APhysPolB.47.2581}{Acta Phys. Polon. B {\bf
  47}, 2581, 2016},
  [\href{http://arxiv.org/abs/arXiv:1610.02023}{{arXiv:1610.02023 [hep-th]}}].

\bibitem{Grozdanov:2016vgg}
S.~Grozdanov, N.~Kaplis and A.~O. Starinets, {{From strong to weak coupling in
  holographic models of thermalization}},
  \href{http://dx.doi.org/10.1007/JHEP07(2016)151}{JHEP {\bf 07}, 151, 2016},
  [\href{http://arxiv.org/abs/arXiv:1605.02173}{{arXiv:1605.02173 [hep-th]}}].

\bibitem{Florkowski:2017olj}
W.~Florkowski, M.~P. Heller and M.~Spalinski, {{New theories of relativistic
  hydrodynamics in the LHC era}},
  \href{http://dx.doi.org/10.1088/1361-6633/aaa091}{Rept. Prog. Phys. {\bf 81},
  046001, 2018},
  [\href{http://arxiv.org/abs/arXiv:1707.02282}{{arXiv:1707.02282 [hep-ph]}}].

\bibitem{Casalderrey-Solana:2018rle}
J.~Casalderrey-Solana, S.~Grozdanov and A.~O. Starinets, {{Transport Peak in
  the Thermal Spectral Function of $\mathcal N=4$ Supersymmetric Yang-Mills
  Plasma at Intermediate Coupling}},
  \href{http://dx.doi.org/10.1103/PhysRevLett.121.191603}{Phys. Rev. Lett. {\bf
  121}, 191603, 2018},
  [\href{http://arxiv.org/abs/arXiv:1806.10997}{{arXiv:1806.10997 [hep-th]}}].

\bibitem{Grozdanov:2018gfx}
S.~Grozdanov and A.~O. Starinets, {{Adding new branches to the
  \textquotedblleft{}Christmas tree\textquotedblright{} of the quasinormal
  spectrum of black branes}},
  \href{http://dx.doi.org/10.1007/JHEP04(2019)080}{JHEP {\bf 04}, 080, 2019},
  [\href{http://arxiv.org/abs/arXiv:1812.09288}{{arXiv:1812.09288 [hep-th]}}].

\bibitem{Grozdanov:2019kge}
S.~Grozdanov, P.~K. Kovtun, A.~O. Starinets and P.~Tadi\'c, {{Convergence of
  the Gradient Expansion in Hydrodynamics}},
  \href{http://dx.doi.org/10.1103/PhysRevLett.122.251601}{Phys. Rev. Lett. {\bf
  122}, 251601, 2019},
  [\href{http://arxiv.org/abs/arXiv:1904.01018}{{arXiv:1904.01018 [hep-th]}}].

\bibitem{Grozdanov:2019uhi}
S.~Grozdanov, P.~K. Kovtun, A.~O. Starinets and P.~Tadi\'c, {{The complex life
  of hydrodynamic modes}},
  \href{http://dx.doi.org/10.1007/JHEP11(2019)097}{JHEP {\bf 11}, 097, 2019},
  [\href{http://arxiv.org/abs/arXiv:1904.12862}{{arXiv:1904.12862 [hep-th]}}].

\bibitem{Berges:2020fwq}
J.~Berges, M.~P. Heller, A.~Mazeliauskas and R.~Venugopalan, {{QCD
  thermalization: Ab initio approaches and interdisciplinary connections}},
  \href{http://dx.doi.org/10.1103/RevModPhys.93.035003}{Rev. Mod. Phys. {\bf
  93}, 035003, 2021},
  [\href{http://arxiv.org/abs/arXiv:2005.12299}{{arXiv:2005.12299 [hep-th]}}].

\bibitem{Heller:2020hnq}
M.~P. Heller, A.~Serantes, M.~Spali\'nski, V.~Svensson and B.~Withers,
  {{Convergence of hydrodynamic modes: insights from kinetic theory and
  holography}}, \href{http://dx.doi.org/10.21468/SciPostPhys.10.6.123}{SciPost
  Phys. {\bf 10}, 123, 2021},
  [\href{http://arxiv.org/abs/arXiv:2012.15393}{{arXiv:2012.15393 [hep-th]}}].

\bibitem{Grozdanov:2021gzh}
S.~Grozdanov, A.~O. Starinets and P.~Tadi\'c, {{Hydrodynamic dispersion
  relations at finite coupling}},
  \href{http://dx.doi.org/10.1007/JHEP06(2021)180}{JHEP {\bf 06}, 180, 2021},
  [\href{http://arxiv.org/abs/arXiv:2104.11035}{{arXiv:2104.11035 [hep-th]}}].

\bibitem{Ghiglieri:2020dpq}
J.~Ghiglieri, A.~Kurkela, M.~Strickland and A.~Vuorinen, {{Perturbative Thermal
  QCD: Formalism and Applications}},
  \href{http://dx.doi.org/10.1016/j.physrep.2020.07.004}{Phys. Rept. {\bf 880},
  1--73, 2020}, [\href{http://arxiv.org/abs/arXiv:2002.10188}{{arXiv:2002.10188
  [hep-ph]}}].

\bibitem{Milhano:2015mng}
J.~G. Milhano and K.~C. Zapp, {{Origins of the di-jet asymmetry in heavy ion
  collisions}}, \href{http://dx.doi.org/10.1140/epjc/s10052-016-4130-9}{Eur.
  Phys. J. C {\bf 76}, 288, 2016},
  [\href{http://arxiv.org/abs/arXiv:1512.08107}{{arXiv:1512.08107 [hep-ph]}}].

\bibitem{Rajagopal:2016uip}
K.~Rajagopal, A.~V. Sadofyev and W.~van~der Schee, {{Evolution of the jet
  opening angle distribution in holographic plasma}},
  \href{http://dx.doi.org/10.1103/PhysRevLett.116.211603}{Phys. Rev. Lett. {\bf
  116}, 211603, 2016},
  [\href{http://arxiv.org/abs/arXiv:1602.04187}{{arXiv:1602.04187 [nucl-th]}}].

\bibitem{Brewer:2017fqy}
J.~Brewer, K.~Rajagopal, A.~Sadofyev and W.~Van Der~Schee, {{Evolution of the
  Mean Jet Shape and Dijet Asymmetry Distribution of an Ensemble of Holographic
  Jets in Strongly Coupled Plasma}},
  \href{http://dx.doi.org/10.1007/JHEP02(2018)015}{JHEP {\bf 02}, 015, 2018},
  [\href{http://arxiv.org/abs/arXiv:1710.03237}{{arXiv:1710.03237 [nucl-th]}}].

\bibitem{Brewer:2021hmh}
J.~Brewer, Q.~Brodsky and K.~Rajagopal, {{Disentangling jet modification in jet
  simulations and in Z+jet data}},
  \href{http://dx.doi.org/10.1007/JHEP02(2022)175}{JHEP {\bf 02}, 175, 2022},
  [\href{http://arxiv.org/abs/arXiv:2110.13159}{{arXiv:2110.13159 [hep-ph]}}].

\bibitem{Binder:2021otw}
T.~Binder, K.~Mukaida, B.~Scheihing-Hitschfeld and X.~Yao, {{Non-Abelian
  electric field correlator at NLO for dark matter relic abundance and
  quarkonium transport}}, \href{http://dx.doi.org/10.1007/JHEP01(2022)137}{JHEP
  {\bf 01}, 137, 2022},
  [\href{http://arxiv.org/abs/arXiv:2107.03945}{{arXiv:2107.03945 [hep-ph]}}].

\bibitem{Yao:2021lus}
X.~Yao, {{Open quantum systems for quarkonia}},
  \href{http://dx.doi.org/10.1142/S0217751X21300106}{Int. J. Mod. Phys. A {\bf
  36}, 2130010, 2021},
  [\href{http://arxiv.org/abs/arXiv:2102.01736}{{arXiv:2102.01736 [hep-ph]}}].

\bibitem{Annala:2017llu}
E.~Annala, T.~Gorda, A.~Kurkela and A.~Vuorinen, {{Gravitational-wave
  constraints on the neutron-star-matter Equation of State}},
  \href{http://dx.doi.org/10.1103/PhysRevLett.120.172703}{Phys. Rev. Lett. {\bf
  120}, 172703, 2018},
  [\href{http://arxiv.org/abs/arXiv:1711.02644}{{arXiv:1711.02644
  [astro-ph.HE]}}].

\bibitem{Annala:2019puf}
E.~Annala, T.~Gorda, A.~Kurkela, J.~N\"attil\"a and A.~Vuorinen, {{Evidence for
  quark-matter cores in massive neutron stars}},
  \href{http://dx.doi.org/10.1038/s41567-020-0914-9}{Nature Phys. {\bf 16},
  907--910, 2020},
  [\href{http://arxiv.org/abs/arXiv:1903.09121}{{arXiv:1903.09121
  [astro-ph.HE]}}].

\bibitem{Annala:2021gom}
E.~Annala, T.~Gorda, E.~Katerini, A.~Kurkela, J.~N\"attil\"a, V.~Paschalidis
  and A.~Vuorinen, {{Multimessenger constraints for ultra-dense matter}},
  [\href{http://arxiv.org/abs/arXiv:2105.05132}{{arXiv:2105.05132
  [astro-ph.HE]}}].

\bibitem{Attems:2019yqn}
M.~Attems, Y.~Bea, J.~Casalderrey-Solana, D.~Mateos and M.~Zilh\~ao, {{Dynamics
  of Phase Separation from Holography}},
  \href{http://dx.doi.org/10.1007/JHEP01(2020)106}{JHEP {\bf 01}, 106, 2020},
  [\href{http://arxiv.org/abs/arXiv:1905.12544}{{arXiv:1905.12544 [hep-th]}}].

\bibitem{Janik:2021jbq}
R.~A. Janik, M.~Jarvinen and J.~Sonnenschein, {{A simple description of
  holographic domain walls in confining theories \textemdash{} extended
  hydrodynamics}}, \href{http://dx.doi.org/10.1007/JHEP09(2021)129}{JHEP {\bf
  09}, 129, 2021},
  [\href{http://arxiv.org/abs/arXiv:2106.02642}{{arXiv:2106.02642 [hep-th]}}].

\bibitem{Bea:2021ieq}
Y.~Bea, J.~Casalderrey-Solana, T.~Giannakopoulos, D.~Mateos,
  M.~Sanchez-Garitaonandia and M.~Zilh\~ao, {{Domain Collisions}},
  [\href{http://arxiv.org/abs/arXiv:2111.03355}{{arXiv:2111.03355 [hep-th]}}].

\bibitem{Annala:2017tqz}
E.~Annala, C.~Ecker, C.~Hoyos, N.~Jokela, D.~Rodr\'\i{}guez~Fern\'andez and
  A.~Vuorinen, {{Holographic compact stars meet gravitational wave
  constraints}}, \href{http://dx.doi.org/10.1007/JHEP12(2018)078}{JHEP {\bf
  12}, 078, 2018},
  [\href{http://arxiv.org/abs/arXiv:1711.06244}{{arXiv:1711.06244
  [astro-ph.HE]}}].

\bibitem{Faedo:2018fjw}
A.~F. Faedo, D.~Mateos, C.~Pantelidou and J.~Tarr\'\i{}o, {{A Supersymmetric
  Color Superconductor from Holography}},
  \href{http://dx.doi.org/10.1007/JHEP05(2019)106}{JHEP {\bf 05}, 106, 2019},
  [\href{http://arxiv.org/abs/arXiv:1807.09712}{{arXiv:1807.09712 [hep-th]}}].

\bibitem{Henriksson:2019zph}
O.~Henriksson, C.~Hoyos and N.~Jokela, {{Novel color superconducting phases of
  $\cal{N}$ = 4 super Yang-Mills at strong coupling}},
  \href{http://dx.doi.org/10.1007/JHEP09(2019)088}{JHEP {\bf 09}, 088, 2019},
  [\href{http://arxiv.org/abs/arXiv:1907.01562}{{arXiv:1907.01562 [hep-th]}}].

\bibitem{Faedo:2019jlp}
A.~F. Faedo, D.~Mateos, C.~Pantelidou and J.~Tarr\'\i{}o, {{Spectrum of a
  Supersymmetric Color Superconductor}},
  \href{http://dx.doi.org/10.1007/JHEP11(2019)020}{JHEP {\bf 11}, 020, 2019},
  [\href{http://arxiv.org/abs/arXiv:1909.00227}{{arXiv:1909.00227 [hep-th]}}].

\bibitem{Hoyos:2020hmq}
C.~Hoyos, N.~Jokela, M.~Jarvinen, J.~G. Subils, J.~Tarrio and A.~Vuorinen,
  {{Transport in strongly coupled quark matter}},
  \href{http://dx.doi.org/10.1103/PhysRevLett.125.241601}{Phys. Rev. Lett. {\bf
  125}, 241601, 2020},
  [\href{http://arxiv.org/abs/arXiv:2005.14205}{{arXiv:2005.14205 [hep-th]}}].

\bibitem{Jokela:2021vwy}
N.~Jokela, M.~J\"arvinen and J.~Remes, {{Holographic QCD in the NICER era}},
  [\href{http://arxiv.org/abs/arXiv:2111.12101}{{arXiv:2111.12101 [hep-ph]}}].

\bibitem{Hoyos:2021uff}
C.~Hoyos, N.~Jokela and A.~Vuorinen, {{Holographic approach to compact stars
  and their binary mergers}},
  [\href{http://arxiv.org/abs/arXiv:2112.08422}{{arXiv:2112.08422 [hep-th]}}].

\bibitem{Hubeny:2010ry}
V.~E. Hubeny and M.~Rangamani, {{A Holographic view on physics out of
  equilibrium}}, \href{http://dx.doi.org/10.1155/2010/297916}{Adv. High Energy
  Phys. {\bf 2010}, 297916, 2010},
  [\href{http://arxiv.org/abs/arXiv:1006.3675}{{arXiv:1006.3675 [hep-th]}}].

\bibitem{Liu:2018crr}
H.~Liu and J.~Sonner, {{Holographic systems far from equilibrium: a review}},
  [\href{http://arxiv.org/abs/arXiv:1810.02367}{{arXiv:1810.02367 [hep-th]}}].

\bibitem{Bhattacharyya:2008jc}
S.~Bhattacharyya, V.~E. Hubeny, S.~Minwalla and M.~Rangamani, {{Nonlinear Fluid
  Dynamics from Gravity}},
  \href{http://dx.doi.org/10.1088/1126-6708/2008/02/045}{JHEP {\bf 0802}, 045,
  2008}, [\href{http://arxiv.org/abs/arXiv:0712.2456}{{arXiv:0712.2456
  [hep-th]}}].

\bibitem{Erdmenger:2008rm}
J.~Erdmenger, M.~Haack, M.~Kaminski and A.~Yarom, {{Fluid dynamics of R-charged
  black holes}}, \href{http://dx.doi.org/10.1088/1126-6708/2009/01/055}{JHEP
  {\bf 0901}, 055, 2009},
  [\href{http://arxiv.org/abs/arXiv:0809.2488}{{arXiv:0809.2488 [hep-th]}}].

\bibitem{Banerjee:2008th}
N.~Banerjee, J.~Bhattacharya, S.~Bhattacharyya, S.~Dutta, R.~Loganayagam
  et~al., {{Hydrodynamics from charged black branes}},
  \href{http://dx.doi.org/10.1007/JHEP01(2011)094}{JHEP {\bf 1101}, 094, 2011},
  [\href{http://arxiv.org/abs/arXiv:0809.2596}{{arXiv:0809.2596 [hep-th]}}].

\bibitem{Son:2009tf}
D.~T. Son and P.~Surowka, {{Hydrodynamics with Triangle Anomalies}},
  \href{http://dx.doi.org/10.1103/PhysRevLett.103.191601}{Phys. Rev. Lett. {\bf
  103}, 191601, 2009},
  [\href{http://arxiv.org/abs/arXiv:0906.5044}{{arXiv:0906.5044 [hep-th]}}].

\bibitem{Heller:2012km}
M.~P. Heller, D.~Mateos, W.~van~der Schee and D.~Trancanelli, {{Strong Coupling
  Isotropization of Non-Abelian Plasmas Simplified}},
  \href{http://dx.doi.org/10.1103/PhysRevLett.108.191601}{Phys. Rev. Lett. {\bf
  108}, 191601, 2012},
  [\href{http://arxiv.org/abs/arXiv:1202.0981}{{arXiv:1202.0981 [hep-th]}}].

\bibitem{Liu:2013iza}
H.~Liu and S.~J. Suh, {{Entanglement Tsunami: Universal Scaling in Holographic
  Thermalization}},
  \href{http://dx.doi.org/10.1103/PhysRevLett.112.011601}{Phys. Rev. Lett. {\bf
  112}, 011601, 2014},
  [\href{http://arxiv.org/abs/arXiv:1305.7244}{{arXiv:1305.7244 [hep-th]}}].

\bibitem{Adams:2013vsa}
A.~Adams, P.~M. Chesler and H.~Liu, {{Holographic turbulence}},
  \href{http://dx.doi.org/10.1103/PhysRevLett.112.151602}{Phys. Rev. Lett. {\bf
  112}, 151602, 2014},
  [\href{http://arxiv.org/abs/arXiv:1307.7267}{{arXiv:1307.7267 [hep-th]}}].

\bibitem{Adams:2012pj}
A.~Adams, P.~M. Chesler and H.~Liu, {{Holographic Vortex Liquids and Superfluid
  Turbulence}}, \href{http://dx.doi.org/10.1126/science.1233529}{Science {\bf
  341}, 368--372, 2013},
  [\href{http://arxiv.org/abs/arXiv:1212.0281}{{arXiv:1212.0281 [hep-th]}}].

\bibitem{Bousso:2022ntt}
R.~Bousso, X.~Dong, N.~Engelhardt, T.~Faulkner, T.~Hartman, S.~H. Shenker and
  D.~Stanford, {{Snowmass White Paper: Quantum Aspects of Black Holes and the
  Emergence of Spacetime}},
  [\href{http://arxiv.org/abs/arXiv:2201.03096}{{arXiv:2201.03096 [hep-th]}}].

\bibitem{PhysRevX.5.041041}
J.~P.~F. {LeBlanc}, A.~E. {Antipov}, F.~{Becca}, I.~W. {Bulik}, G.~K.-L.
  {Chan}, C.-M. {Chung}, Y.~{Deng}, M.~{Ferrero}, T.~M. {Henderson}, C.~A.
  {Jim{\'e}nez-Hoyos}, E.~{Kozik}, X.-W. {Liu}, A.~J. {Millis} et~al.,
  {{Solutions of the Two-Dimensional Hubbard Model: Benchmarks and Results from
  a Wide Range of Numerical Algorithms}},
  \href{http://dx.doi.org/10.1103/PhysRevX.5.041041}{Physical Review X {\bf 5},
  041041, 2015},
  [\href{http://arxiv.org/abs/arXiv:1505.02290}{{arXiv:1505.02290
  [cond-mat.str-el]}}].

\bibitem{PhysRevX.7.031059}
M.~{Motta}, D.~M. {Ceperley}, G.~K.-L. {Chan}, J.~A. {Gomez}, E.~{Gull},
  S.~{Guo}, C.~A. {Jim{\'e}nez-Hoyos}, T.~N. {Lan}, J.~{Li}, F.~{Ma}, A.~J.
  {Millis}, N.~V. {Prokof'ev}, U.~{Ray} et~al., {{Towards the Solution of the
  Many-Electron Problem in Real Materials: Equation of State of the Hydrogen
  Chain with State-of-the-Art Many-Body Methods}},
  \href{http://dx.doi.org/10.1103/PhysRevX.7.031059}{Physical Review X {\bf 7},
  031059, 2017},
  [\href{http://arxiv.org/abs/arXiv:1705.01608}{{arXiv:1705.01608
  [physics.comp-ph]}}].

\bibitem{Bousso:2000xa}
R.~Bousso and J.~Polchinski, {{Quantization of four form fluxes and dynamical
  neutralization of the cosmological constant}},
  \href{http://dx.doi.org/10.1088/1126-6708/2000/06/006}{JHEP {\bf 06}, 006,
  2000},
  [\href{http://arxiv.org/abs/arXiv:hep-th/0004134}{{arXiv:hep-th/0004134}}].

\bibitem{Denef:2011ee}
F.~Denef, {{TASI lectures on complex structures}},  in \emph{{Theoretical
  Advanced Study Institute in Elementary Particle Physics}: {String theory and
  its Applications: From meV to the Planck Scale}}, pp.~407--512, 2011.
\newblock [\href{http://arxiv.org/abs/arXiv:1104.0254}{{arXiv:1104.0254
  [hep-th]}}].

\bibitem{Dong:2010pm}
X.~Dong, B.~Horn, E.~Silverstein and G.~Torroba, {{Micromanaging de Sitter
  holography}}, \href{http://dx.doi.org/10.1088/0264-9381/27/24/245020}{Class.
  Quant. Grav. {\bf 27}, 245020, 2010},
  [\href{http://arxiv.org/abs/arXiv:1005.5403}{{arXiv:1005.5403 [hep-th]}}].

\bibitem{Anninos:2012qw}
D.~Anninos, {{De Sitter Musings}},
  \href{http://dx.doi.org/10.1142/S0217751X1230013X}{Int. J. Mod. Phys. A {\bf
  27}, 1230013, 2012},
  [\href{http://arxiv.org/abs/arXiv:1205.3855}{{arXiv:1205.3855 [hep-th]}}].

\bibitem{Anninos:2020hfj}
D.~Anninos, F.~Denef, Y.~T.~A. Law and Z.~Sun, {{Quantum de Sitter horizon
  entropy from quasicanonical bulk, edge, sphere and topological string
  partition functions}}, \href{http://dx.doi.org/10.1007/JHEP01(2022)088}{JHEP
  {\bf 01}, 088, 2022},
  [\href{http://arxiv.org/abs/arXiv:2009.12464}{{arXiv:2009.12464 [hep-th]}}].

\bibitem{PhysRevX.10.011069}
X.~Han and S.~A. Hartnoll, {{Deep Quantum Geometry of Matrices}},
  \href{http://dx.doi.org/10.1103/PhysRevX.10.011069}{Phys. Rev. X {\bf 10},
  011069, 2020},
  [\href{http://arxiv.org/abs/arXiv:1906.08781}{{arXiv:1906.08781 [hep-th]}}].

\bibitem{Lin:2020mme}
H.~W. Lin, {{Bootstraps to strings: solving random matrix models with
  positivity}}, \href{http://dx.doi.org/10.1007/JHEP06(2020)090}{JHEP {\bf 06},
  090, 2020}, [\href{http://arxiv.org/abs/arXiv:2002.08387}{{arXiv:2002.08387
  [hep-th]}}].

\bibitem{Han:2020bkb}
X.~Han, S.~A. Hartnoll and J.~Kruthoff, {{Bootstrapping Matrix Quantum
  Mechanics}}, \href{http://dx.doi.org/10.1103/PhysRevLett.125.041601}{Phys.
  Rev. Lett. {\bf 125}, 041601, 2020},
  [\href{http://arxiv.org/abs/arXiv:2004.10212}{{arXiv:2004.10212 [hep-th]}}].

\bibitem{han2020quantum}
X.~Han, {Quantum many-body bootstrap},
  [\href{http://arxiv.org/abs/arXiv:2006.06002}{{arXiv:2006.06002
  [cond-mat.str-el]}}].

\bibitem{Frenkel:2020ysx}
A.~Frenkel, S.~A. Hartnoll, J.~Kruthoff and Z.~D. Shi, {{Holographic flows from
  CFT to the Kasner universe}},
  \href{http://dx.doi.org/10.1007/JHEP08(2020)003}{JHEP {\bf 08}, 003, 2020},
  [\href{http://arxiv.org/abs/arXiv:2004.01192}{{arXiv:2004.01192 [hep-th]}}].

\bibitem{Hartnoll:2020rwq}
S.~A. Hartnoll, G.~T. Horowitz, J.~Kruthoff and J.~E. Santos, {{Gravitational
  duals to the grand canonical ensemble abhor Cauchy horizons}},
  \href{http://dx.doi.org/10.1007/JHEP10(2020)102}{JHEP {\bf 10}, 102, 2020},
  [\href{http://arxiv.org/abs/arXiv:2006.10056}{{arXiv:2006.10056 [hep-th]}}].

\bibitem{Hartnoll:2020fhc}
S.~A. Hartnoll, G.~T. Horowitz, J.~Kruthoff and J.~E. Santos, {{Diving into a
  holographic superconductor}},
  \href{http://dx.doi.org/10.21468/SciPostPhys.10.1.009}{SciPost Phys. {\bf
  10}, 009, 2021},
  [\href{http://arxiv.org/abs/arXiv:2008.12786}{{arXiv:2008.12786 [hep-th]}}].

\bibitem{Leutheusser:2021qhd}
S.~Leutheusser and H.~Liu, {{Causal connectability between quantum systems and
  the black hole interior in holographic duality}},
  [\href{http://arxiv.org/abs/arXiv:2110.05497}{{arXiv:2110.05497 [hep-th]}}].

\bibitem{Leutheusser:2021frk}
S.~Leutheusser and H.~Liu, {{Emergent times in holographic duality}},
  [\href{http://arxiv.org/abs/arXiv:2112.12156}{{arXiv:2112.12156 [hep-th]}}].

\end{thebibliography}\endgroup

\end{document}